\documentclass[authoryear,12pt,round]{article}

\RequirePackage{amsthm,amsmath,amsfonts,amssymb}
\RequirePackage[authoryear]{natbib}
\RequirePackage[colorlinks,citecolor=blue,urlcolor=blue]{hyperref}
\RequirePackage{graphicx}
\RequirePackage{nameref}
\RequirePackage{bm}
\RequirePackage{float}
\RequirePackage{color,grffile}
\RequirePackage{times,subcaption}
\RequirePackage{rotating}
\RequirePackage{mathtools}
\RequirePackage[plain,noend]{algorithm2e}

\graphicspath{ {Figures/} }

\makeatletter
\let\orgdescriptionlabel\descriptionlabel
\renewcommand*{\descriptionlabel}[1]{%
  \let\orglabel\label
  \let\label\@gobble
  \phantomsection
  \edef\@currentlabel{#1}%
  \let\label\orglabel
  \orgdescriptionlabel{#1}%
}
\makeatother

\newcommand{\Nbiter}{{L}}
\newcommand{\nbiter}{{\ell}}
\newcommand{\Nbsim}{{S}}
\newcommand{\Nbvar}{{p}}
\newcommand{\bfx}{{\mathbf{x}}}

\newcommand{\bfz}{{\mathbf{z}}}
\newcommand{\bfZ}{{\mathbf{Z}}}
\newcommand{\bfW}{{\mathbf{W}}}
\newcommand{\bfw}{{\mathbf{w}}}
\newcommand{\bfR}{{\mathbf{R}}}
\newcommand{\bfM}{{\mathbf{M}}}
\newcommand{\bfA}{{\mathbf{A}}}
\newcommand{\bfB}{{\mathbf{B}}}
\newcommand{\bfH}{{\mathbf{H}}}
\newcommand{\bfU}{{\mathbf{U}}}
\newcommand{\bfu}{{\mathbf{u}}}
\newcommand{\nbvar}{{j}}
\newcommand{\Nbgroup}{{K}}
\newcommand{\nbgroup}{{w}}
\newcommand{\intra}{{V}}
\newcommand{\Nbind}{{n}}
\newcommand{\nbind}{{i}}

\newcommand{\Nbtab}{{M}}
\newcommand{\nbtab}{{m}}

\newcommand{\bfbeta}{{\bm{\beta}}}
\newcommand{\bfmu}{{\bm{\mu}}}
\newcommand{\bftheta}{{\bm{\theta}}}
\newcommand{\bfSigma}{{\bm{\Sigma}}}
\newcommand{\bfzeta}{{\bm{\zeta}}}
\newcommand{\bfalpha}{{\bm{\alpha}}}

\title{Clustering with missing data: which imputation model for which cluster analysis method?}
\author{Vincent Audigier\footnote{CNAM, Laboratoire Cedric-MSDMA, 2 rue Cont\'e, 75003 Paris, France (e-mail: vincent.audigier@cnam.fr, n-deye.niang\_ keita@cnam.fr)}, Nd\`eye Niang\footnotemark[1], Matthieu Resche-Rigon\footnote{Service de Biostatistique et Information M\'edicale, H\^opital
Saint-Louis, AP-HP, Paris, France; Universit\'e de Paris, Paris, France; INSERM, UMR 1153, Equipe ECSTRRA, Paris,
France (e-mail: matthieu.resche-rigon@u-paris.fr)}
}

\begin{document}
\maketitle

\textbf{Abstract} Multiple imputation (MI) is a popular method for dealing with missing values. One main advantage of MI is to separate the imputation phase and the analysis one. However, both are related since they are based on distribution assumptions that have to be consistent. This point is well known as \textit{congeniality}.

In this paper, we discuss congeniality for clustering on continuous data. First, we theoretically highlight how two joint modeling (JM) MI methods (JM-GL and JM-DP) 
are congenial with various clustering methods. Then, we propose a new fully conditional specification (FCS) MI method with the same theoretical properties as JM-GL. Finally, we extend this FCS MI method to account for more complex distributions. Based on an extensive simulation study, all MI methods are compared for various cluster analysis methods (k-means, k-medoids, mixture model, hierarchical clustering).

This study highlights the partition accuracy is improved when the imputation model accounts for clustered individuals. From this point of view, standard MI methods ignoring such a structure should be avoided. JM-GL and JM-DP should be recommended when data are distributed according to a gaussian mixture model, while FCS methods outperform JM ones on more complex data.

\vspace{1cm}

\section{Introduction}
Many statistical methods have been proposed for identifying clusters of individuals in data sets (see \cite{Jain88} for a review). Two categories of methods are often distinguished: the model-based methods, which explicitly assume a statistical model accounting for the structure in clusters; and the distance-based methods, which define a distance (or simply a dissimilarity) between individuals and then gather closest individuals in a common cluster (like $k-$means clustering \citep{Forgy65}, Partitioning Around Medoids (pam) \citep{Kaufman90}, or hierarchical clustering \citep{Ward63}). 

Missing values are a pregnant issue in clustering, which cannot be easily addressed. Indeed, removing incomplete observations would imply incomplete individuals could not be clustered, while removing incomplete variables would require having complete variables. Furthermore, there is no guarantee these variables can be used to identify clusters. Thus, more sophisticated methods have been proposed by various authors. Among model-based methods, \cite{Marbac19} proposed to use mixture models to perform clustering on high dimensional mixed data, while \cite{Miao16,Marbac20} studied mixture models for addressing MNAR mechanisms for instance.

Among distance-based methods, \cite{Chi16, Konda11,Wagstaff04} extended k-means clustering to handle missing values, while \cite{Zhang2016fuzzy,Hathaway01} adapted fuzzy C-means clustering to missing values.

While multiple imputation is a very popular method to handle missing values \citep{Rubin76, Rubin87}, such a technique has not been so considered in the literature of clustering with missing values. Two main reasons can be raised. First, the advantage to use multiple imputation (MI) over single imputation (SI) was unclear for cluster analysis. Then, until recently the rules to pool the partitions after MI remained unclear. Indeed, compared to SI, MI is motivated by assessing the sampling variability by accounting for missing values. However, previous works in clustering only focus the way to aggregate the partitions. \cite{Plaehn19} proposed to use stacking methods, while \cite{Faucheux,Bruckers17,Basagana13} proposed to use various consensus clustering methods. \cite{Audigier20} filled the gap by proposing an equivalent for Rubin's Rules in clustering. Furthermore, they demonstrated MI improves accuracy compared to SI, in the spirit of bagging clustering \citep{Dudoit03}.

A main advantage of dealing missing values by MI is that it separates the imputation step and the analysis step. In particular, it allows applying any clustering method even if this method cannot be modified to handle missing values. However, this distinctness between imputation and analysis also brings the \textit{uncongeniality} \citep{Meng94,Schafer03} issue. Roughly speaking, congeniality refers to the adequacy between the model used by the imputer and the one used by the analyst. Since uncongenial imputation can bias the analysis results \citep{Seaman16}, the choice of the imputation model is a main issue in order to be able to consider MI as a solution to the problem of missing values in clustering.

Two ways are generally distinguished for imputing data: joint modeling (JM) MI \citep{Schafer97} and fully conditional specification (FCS) MI \citep{vanBuuren06}. JM MI consists in explicitly defining a joint distribution for all variables, while FCS MI consists in defining one model per incomplete variable (under some conditions, it also corresponds to defining implicitly a joint distribution). 

JM MI is theoretically appealing, but FCS MI can lead to more efficient statistical inferences when the imputation model is miss-specified. Moreover, even if the model is well specified, the loss by using FCS MI remains small \citep{Seaman16}. Furthermore, by tuning conditional distributions of the imputation model, FCS MI can be easily modified to account for specific data structure, while JM MI remains difficult to modify \citep{VB18}.

Thus, the objectives of this paper consist in highlighting which MI method can be congenial for clustering under a missing at random mechanism \citep{Rubin76} and to propose a new congenial FCS MI method for achieving this goal.

In the next section, the statistical context is outlined (Section \ref{Section1}). Then, congeniality is discussed for two JM MI methods in Section \ref{Section2} and a new FCS MI method is proposed in Section \ref{Section3}. Finally, MI methods are compared by an extensive simulation study in Section \ref{Section4} and a discussion closes the paper (Section \ref{Section5}).

\section{Notations and context\label{Section1}}
\subsection{Notations}
Let $\bfZ=\left(z_{\nbind\nbvar}\right)_{\begin{array}{c}
    1\leq \nbind\leq\Nbind   \\
    1\leq \nbvar\leq\Nbvar  
\end{array}}$ be a data set. We assume each individual $\nbind$ ($1\leq\nbind\leq\Nbind$) belongs to a unique cluster $\nbgroup\in \{1,...,\Nbgroup\}$. Based on individual profiles $(\bfz_\nbind)_{(1\leq\nbind\leq\Nbind)}$, we aim at identifying the cluster (denoted $\nbgroup_\nbind$) of each individual . Let $W$ be the categorical variable with $\Nbgroup$ categories associated to the cluster assignment and  $Z=\left(Z_1,..,Z_\Nbvar\right)$ be the set of random continuous variables associated to $\bfZ$. Distributions of random variables are denoted by $p(.)$. Additionally, we assume some values of $\bfZ$ are missing. Let $\bfR=\left(r_{\nbind\nbvar}\right)_{\begin{array}{c}
    1\leq \nbind\leq\Nbind   \\
    1\leq \nbvar\leq\Nbvar  
\end{array}}$ be the missing data pattern so that $r_{\nbind\nbvar}=0$ if $z_{\nbind\nbvar}$ is missing and $r_{\nbind\nbvar}=1$ otherwise. The missing part of $\bfz$ (a realization of $Z$) is denoted $\bfz^{miss}$ and the observed part $\bfz^{obs}$. The associated random variables are denoted $R$, $Z^{miss}$ and $Z^{obs}$ respectively. We place ourselves under the missing at random assumption, meaning $R$ and $Z^{miss}$ are independent conditionally to $Z^{obs}$.

\subsection{Clustering on full data\label{clustsection}}

Model-based clustering is classically based on finite mixture models \citep{mclachlan88} defined as follows:

\begin{eqnarray}
    W\sim \mathcal{M}\left(1,\bftheta\right)\\
    p(Z=z)=\sum_{\nbgroup=1}^\Nbgroup p\left(W=\nbgroup\right)f_\nbgroup(z)
\end{eqnarray}
where $\mathcal{M}$ indicates a multinomial distribution, $\bftheta=\left(\theta_1,...,\theta_\Nbgroup\right)^t$ the vector of probabilities associated to each category (\textit{i.e.} each cluster) and $f_\nbgroup(z)$ denotes the density of the $\nbgroup$th component.

Mixture model generally assumes normality for $f_\nbgroup$ and so are called gaussian mixture model. The parameters of this model are $\bftheta$, as well as $\bfmu=\left(\bfmu_1,...,\bfmu_\Nbgroup\right)$, \textit{i.e.} the mean vectors of each mixture component, and $\bfSigma=\left(\bfSigma_1,...,\bfSigma_\Nbgroup\right)$, \textit{i.e.} the covariance matrices of each mixture component. The set of parameters is denoted $\bfzeta=\left(\bftheta,\bfmu,\bfSigma\right)$. Constraints are generally assumed on $\bfzeta$ to limit overfitting. Among them, the homoscedasticity constraint consists in assuming constant variance across clusters (in contrast to the heteroscedasticity assumption). The maximum-likelihood estimate is generally obtained using an EM algorithm \citep{mclachlan88}.

From this model, a realization $z$ can be clustered by looking at
\begin{equation}
    argmax_{\nbgroup\in\{1,...,\Nbgroup\}} f_\nbgroup(z)p\left(W=\nbgroup\right)
\end{equation} corresponding to the class maximizing the posterior probability $p\left(W=\nbgroup\vert Z=z\right)$.

Finite gaussian mixture models are popular and well-studied. In particular, many relationships with other clustering methods have been established. Among model-based method, finite gaussian mixture model is asymptotically equivalent to finite mixture of student \citep{Bouveyron19}, meaning that when the number of degrees of freedom increases, both produce the same clusters (in expectation). Then, among distance-based methods, clustering by finite gaussian mixture models is asymptotically equivalent to k-means clustering when variables are independent and the variance tends toward 0 
 \citep{Hastie09}. In addition, because of the equivalence of pam clustering and kmeans for Euclidean distance 
  \citep{Hastie09}, gaussian mixture model also generalizes pam clustering.
Because of these relationships with other popular clustering methods (distance-based or not) and because congeniality refers to distribution assumptions, in the following, we will focus on cluster analysis by gaussian mixture models.

\subsection{Handling missing values in clustering by MI}

\subsubsection{MI principle}
MI for cluster analysis consists of three steps: i) imputation of missing values according to an imputation model $g_{imp}$ $\Nbtab$ times. Step i) provides $\Nbtab$ data sets $\left(\bfZ^{obs},\bfZ^{miss}_\nbtab\right)_{1\leq\nbtab\leq\Nbtab}$ ii) analysis of the $\Nbtab$ imputed data sets according to a cluster analysis method $g_{ana}$ (\textit{e.g.} a mixture model). For each imputed data set, Step ii) provides a partition $\Psi_\nbtab$ and an associated instability $\intra_\nbtab$ \citep{Fang12} iii) pooling of analysis results \citep{Audigier20}:
\begin{itemize}
    \item the pooled partition is defined as
\begin{equation}
    arg min_{\bfH}\parallel \bfM-\bfH\parallel^2 \label{newrule1bis}
\end{equation}
with $\bfM=\frac{1}{\Nbtab}\sum_{\nbtab=1}^\Nbtab \bfH_\nbtab$  and $\bfH_\nbtab$ the connectivity matrix associated to $\Psi_\nbtab$. The solution of this optimization problem can be obtained using non-negative matrix factorization \citep{Li07}.
\item the instability is assessed as follows
\begin{equation}
T=\frac{1}{\Nbtab}\sum_{\nbtab=1}^\Nbtab \intra_\nbtab+\frac{1}{\Nbtab^2} \sum_{\nbtab=1}^\Nbtab\sum_{\nbtab'=1}^\Nbtab \delta\left(\Psi_\nbtab,\Psi_{\nbtab'}\right)/\Nbind^2
\label{B3} 
\end{equation} 
where $\delta$ corresponds to the Mirkin distance between partitions, \textit{i.e.} the number of disagreements between partitions.
\end{itemize}

\subsubsection{Congeniality \label{sec:cong}}

Congeniality refers to the adequacy between the model assumed for the analysis and the model used for imputation. In our context, we aim at estimating $\bfzeta$ the set of parameters in a gaussian mixture model from incomplete data. For achieving this goal, we would like to use MI through an imputation model $g_{imp}$. The concept of congeniality has first been developed in \cite{Meng94} and can be defined as follows \citep{Murray18,Bartlett20}: the analysis model $g_{ana}$ is congenial to the imputation model $g_{imp}$ if we can find a Bayesian model $g$ so that:
\begin{enumerate}
    \item given imputed data, $\left(\widehat{\bfzeta},\widehat{Var}\left(\widehat{\bfzeta}\right)\right)$ is asymptotically the same as the posterior mean and variance of $\bfzeta$ under $g$ \label{cong1B}
    \item the predictive distribution of missing values under $g_{imp}$ is the same under $g$\label{cong2}.
\end{enumerate}

The congeniality of the imputation model is most of the time difficult to prove, but if both the imputation model and the analysis one model the same set of variables, then a sufficient condition to ensure congeniality is that both models are identical (whatever the prior distribution of parameters). Thus, we subsequently identify imputation models based on gaussian mixture models. 

\section{JM MI for clustering\label{Section2}}

Two JM MI methods based on gaussian mixture models can be highlighted: the imputation using the general location model (JM-GL) proposed in \cite{Schafer97} and imputation based on the Dirichlet process mixture of products of multivariate normal distributions (JM-DP) proposed in \cite{Kim14}.
\subsection{JM-GL}
\subsubsection{Imputation model}
The general location model can be seen as the gold standard for mixed data imputation. For our purpose, we present it in the case where all variables are continuous and only one variable is categorical. Furthermore, in this subsection the model is presented assuming continuous variables are complete. The incomplete case will be treated in the next subsection.

For convenience, the categorical variable is represented by the  vector of absolute frequencies $\bfx=\left(x_1,...,x_\Nbgroup\right)$ or its disjunctive table $\bfU$ $\left(\Nbind\times\Nbgroup\right)$, so that $diag(\bfU^t\bfU)=\bfx$. The general location model is as follows
\begin{eqnarray}
    x\sim \mathcal{M}(\Nbind,\bftheta)\\
    Z\vert W=\nbgroup\sim \mathcal{N}\left(\mu_{\nbgroup},\bfSigma\right)
\end{eqnarray}
The parameters of this model $\bftheta=\left(\theta_1,...,\theta_\Nbgroup\right)^t$ the vector of proportion of each category, $\bfmu=\left(\bfmu_1,...,\bfmu_{\Nbgroup}\right)^t$ the vector of means and $\bfSigma$ the covariance matrix, are denoted $\bfzeta=\left(\bftheta,\bfmu,\bfSigma\right)$.

In this specific case where the data set contains only one categorical variable, the general location model can consequently be seen as a gaussian mixture model where the data set gathers both the continuous variables and the cluster variable.

We can also note this model can be seen as a generalization of linear discriminant analysis for a number of class larger than 2. Furthermore, the distribution of $Z$ given $U$ can also be expressed as a linear model
\begin{equation}
    Z_\nbind=\bfu_\nbind\bfmu+\varepsilon_\nbind~~\text{for all }\left(1\leq\nbind\leq\Nbind\right) \label{regression}
\end{equation}
where the realizations $\varepsilon$ are independent from $\mathcal{N}\left(0,\bfSigma\right)$.
The model can be straightforwardly extended to several categorical variables by replacing all categorical variables by only one taking its categories among the set of combinations \citep{Schafer97}.

A common Bayesian formulation of the general location model consists in considering the following non-informative priors:
\begin{eqnarray}
    \theta\sim \mathcal{D}\left(\bfalpha\right)\\
    \left(\mu,\Sigma\right)\propto \vert\Sigma\vert^{-\frac{\Nbvar+1}{2}}
\end{eqnarray}
with $\mathcal{D}$ the Dirichlet distribution, $\bfalpha$ a $\Nbgroup$-dimensional vector which can be tuned to $x/\Nbind$. Note that both prior distributions are assumed independent. From this modelling, independent posterior distributions are as follows:
\begin{eqnarray}
\theta\vert W \sim \mathcal{D}\left(\bfalpha+\bfx\right) \label{post1}\\
    \mu\vert\Sigma,Z,W\sim \mathcal{N}\left(\widehat{\bfmu},\bfSigma \otimes diag(\bfx)^{-1}\right)\label{post2}\\
    \Sigma\vert Z,W\sim \mathcal{W}^{-1}\left(\Nbind-\Nbvar,\left(\widehat{\epsilon}^t\widehat{\epsilon}\right)^{-1}\right)\label{post3}
\end{eqnarray}
with $\widehat{\bfmu}$ the estimate of model \eqref{regression}, $\widehat{\epsilon}$ the associated residuals, $\otimes$ the Kronecker product and $\mathcal{W}^{-1}$ the inverse Wishart distribution.
\subsubsection{Data-augmentation}
In our context, $W$ is not observed at all and $Z$ is partially observed. In such a case, drawing $\bfzeta$ in its posterior distribution requires a data-augmentation algorithm \citep{Tanner87}. This algorithm is a Gibbs sampler consisting in alternating two steps until convergence: (I) imputation of data given parameters (P) drawing parameters from their posterior distribution.

Applied to the general location model, the data-augmentation algorithm is as follows \citep{Schafer97}:
\begin{enumerate}
    \item initialize $\bfzeta$ by $\bfzeta^{(0)}=\widehat{\bfzeta}_{ML}$, \textit{i.e.} the maximum likelihood estimator obtained by an EM algorithm
    \item for $\nbiter$ in $1,...,\Nbiter$
    \begin{description}
    \item[I-step\label{istep}] for each $\nbind$ in $\{1,...,\Nbind\}$
    \begin{enumerate}
        \item \label{stepjmgl2} draw $w_\nbind^{(\nbiter)}$ from  $p\left(W\vert Z^{obs}=\bfz_\nbind^{obs},\zeta=\bfzeta^{(\nbiter-1)}\right)$
        \item  \label{stepjmgl22}draw ${z_\nbind^{miss}}^{(\nbiter)}$ from $p(Z^{miss}\vert W={\nbgroup}_\nbind^{(\nbiter)},Z^{obs}=\bfz_\nbind^{obs},\mu=\bfmu^{(\nbiter-1)},\Sigma=\bfSigma^{(\nbiter-1)})$
    \end{enumerate}
    \item[P-step\label{pstep}]~
    \begin{enumerate}
        \item \label{stepjmgl1}draw $\theta^{(\nbiter)}$ from $p\left(\theta\vert W=\bfw^{(\nbiter)}\right)$
        \item draw $\Sigma^{(\nbiter)}$ from $p\left(\Sigma\vert \theta=\bftheta^{(\nbiter)},Z=\bfZ^{(\nbiter)}\right)$
        \item draw $\mu^{(\nbiter)}$ from $p\left(\mu\vert \theta=\bftheta^{(\nbiter)},\Sigma=\bfSigma^{(\nbiter)},Z=\bfZ^{(\nbiter)},W=\bfw^{(\nbiter)}\right)$
    \end{enumerate}
    \end{description}
\end{enumerate}

More details are provided in Appendix \ref{DAGL}. In order to perform MI, several iterations are passed through a burn-in period, then $\Nbtab$ values of $\zeta$ are drawn from the stationary posterior distribution. Note that to avoid dependence between successive draws, several iterations are passed between each one. Finally, the data set can be imputed $\Nbtab$ times by considering each generated value of $\zeta$. Such an algorithm is implemented in the R package mix \citep{mixpackage}.

\subsubsection{Congeniality}
Without missing values, the general location model and the gaussian mixture model are very close. However, several differences can be pointed out, making the imputation model potentially uncongenial with the analysis model.
First, the general location model additionally assumes to know the cluster variable, while this variable is unknown in a gaussian mixture model. Second, the general location model assumes a constant variance within each group, while the gaussian mixture model can handle components with various covariance matrices. Third, the imputation model has a Bayesian formulation, while the gaussian mixture model presented in section \ref{clustsection} has not.

As stayed previously, congeniality can be ensured when both models are identical. The first difference can be easily lift by assuming the categorical variable is completely missing in the general location model and using a data-augmentation algorithm for inference. The second can be tackled by fitting a gaussian mixture model with constant covariance matrix. Note that the difference is due to an extra assumption done in the imputation model. If this assumption is supported by the data, this additional assumption is known to be beneficial for the analysis and corresponds to a case of \textit{superefficiency} \citep{Rubin96}, meaning notably the estimator obtained using MI is more efficient than the maximum likelihood estimator. Otherwise, if this assumption is not relevant, JM-GL could lead to biased estimates. The third difference does not imply uncongeniality. Indeed, the maximum-likelihood of $\bfzeta$ is asymptotically equivalent to its the Bayesian estimator when the number of individuals tends to infinity. Thus, both are asymptotically equivalent, which is sufficient to satisfy the first condition under which models are congenial (Section \ref{sec:cong}).

Finally, the general location model is congenial with a gaussian mixture model provided that the gaussian mixture model assumes homoscedasticity. Otherwise, the imputation model could lead to a biased analysis.

\subsection{JM-DP}
\subsubsection{Imputation model}
Contrary to JM-GL, JM-DP is explicitly based on a gaussian mixture model and uses a data augmentation algorithm for drawing parameters and imputed values. The Bayesian formulation of the model is as follows:
\begin{eqnarray*}
\mu_\nbgroup\vert\Sigma_\nbgroup\sim \mathcal{N}\left(\mu_0,h^{-1}\Sigma_\nbgroup\right)\\
\Sigma_\nbgroup\sim \mathcal{W}^{-1}\left(df,G\right)\\
\text{with } G=\left(g_1,...,g_\Nbvar\right) \ g_\nbvar\sim \mathcal{G}\left(a_0,b_0\right)\\
\theta_\nbgroup=v_\nbgroup \prod_{\ell<\nbgroup}\left(1-v_\ell\right)\\\text{with } \left\lbrace \begin{array}{l}
    v_\nbgroup\sim\mathcal{B}eta\left(1,\alpha\right)\text{ and } \alpha\sim\mathcal{G}(a_\alpha,b_\alpha) \text{ for } \nbgroup<\Nbgroup \\
    v_\Nbgroup=1
\end{array}\right.
\end{eqnarray*}
where $\mathcal{G}$ corresponds to the gamma distribution. $h$, $\mu_0$, $df$, $a_0$, $b_0$, $a_\alpha$, $b_\alpha$ are the hyperparameters of the model (see \cite{Kim14} for tuning). MI under this model is implemented in the R package DPImputeCont \citep{Kimpackage}.

A main property of this model is that it is based on a stick breaking process, allowing an automatic choice of the number of non-empty clusters according to the data. Consequently, the number of clusters has only to be upwardly bounded by $\Nbgroup$.

\subsubsection{Congeniality}
The imputation model used in JM-DP is very close to the gaussian mixture model, but several sources of uncongeniality could be underlined. First, the imputation model does not impose any constraint on the covariance matrices, while those constraints are generally considered for fitting a gaussian mixture model. If data support non-constant covariance structure in each cluster, the imputation model is in this case is more precise than the analysis one. Second, the imputation model allows empty clusters, while the gaussian mixture model cannot be fitted with empty clusters.

The first source of uncongeniality can be tackled by only fitting gaussian mixture models with free covariance matrices, but this kind of uncongeniality generally results only in conservative inference without introducing any bias \citep{Schafer97}. About the empty clusters, this issue is not occurred on large sample because the proportion of each cluster is almost surely over than 0. Since the first condition to ensure congeniality is only asymptotic (Section \ref{sec:cong}), this point is not binding.

Finally, JM-DP is congenial with a gaussian mixture model provided
that the gaussian mixture model assumes heteroscedasticity and that the sample size is large. Otherwise, small sample size could lead to a biased analysis, while an homoscedastic gaussian mixture model could lead to a conservative one.

\subsection{Summary}
As explained previously, statistical properties of both JM methods depend on the homoscedastic or heteroscedastic assumption supported by the data (the "true" model), by the analysis model and by the imputation model. These properties are summarized in Table \ref{tab:summaryjm}.

\section{FCS MI for clustering\label{Section3}}
Based on JM-GL, we now propose a congenial FCS imputation for gaussian mixture models. It first assumes constant variance across latent clusters (as the general location model). Then, we explain how it can be straightforwardly extended to account for congeniality, in particular when the analyst assumes heteroscedaticity for the mixture model.

\subsection{Homoscedastic imputation\label{fcs-homo}}
 As JM MI by data-augmentation, FCS MI can be presented as a Gibbs sampler. However, while the data-augmentation algorithm alternates draws of parameters from their posterior distribution and draws of all missing values from their predictive distribution, FCS alternates both steps variable-by-variable. Similarly to JM-GL, we distinguish the imputation of continuous variables knowing the categorical variable $W$ and the update of the categorical variable.

 Given the categorical variable $W$, the sequential imputation of continuous variables is straightforward. Indeed, given the cluster, continuous variables follow a multivariate distribution. As shown in \cite{Hughes14}, the FCS imputation by regression and the joint imputation are equivalent in such a case, meaning \ref{istep}~\ref{stepjmgl22} in JM-GL can be replaced by sequential imputations.
 
 The imputation of the categorical variable given the continuous ones is more challenging for two reasons. First, the posterior distribution of $\bftheta$ depends on $W$ (\ref{pstep}~\ref{stepjmgl1} in JM-GL) which is unknown at the current step since the goal is to impute this variable. Second, the predictive distribution of missing values depends on $\bfmu$ and $\bfSigma$ (\ref{istep}~ \ref{stepjmgl2} in JM-GL) which are also not available during imputation of the categorical variable since both are specific to a joint distribution. To overcome both difficulties, knowing only continuous variables $\bfZ$ we propose to generate the joint distribution of $(\theta,W)$ at iteration $\nbiter$ as follows:  


\begin{enumerate}
\item \label{fcshomogen1}Drawing $\bftheta^{(\nbiter)}$ from $p(\bftheta\vert Z)$ \begin{enumerate}
\item \label{fcshomogen1a}perform a non-parametric bootstrap on $\bfZ^{(\nbiter)}$:  sample the rows of $\bfZ^{(\nbiter)}$ with replacement one time. Estimate $\bfzeta^\star$ (the parameter of a homoscedastic gaussian mixture model) using an EM algorithm 
    \item \label{fcshomogen1b}draw $W$ from $p(W\vert Z,\bfzeta=\bfzeta^\star)$
    \item \label{fcshomogen1c}generate $\bftheta^{(\nbiter)}$ from $p(\bftheta\vert W,Z)$
    \end{enumerate}
    \item \label{fcshomogen2}Drawing $W^{(\nbiter)}$ from $p(W\vert Z,\bftheta^{(\nbiter)},\bfmu^\star,\Sigma^\star)$
\end{enumerate}

Steps \ref{fcshomogen1a} and \ref{fcshomogen1b} consists in drawing $W$ from $p(W\vert Z)$ by taking advantage from the following hierarchical Bayesian structure \citep{Robert07} \begin{equation*}
    p(W\vert Z) =\int p(W\vert Z,\zeta) p\left(\zeta\vert Z\right)d \zeta
\end{equation*}. Note that the non-parametric bootstrap avoid improper imputation by considering generated values for ($\bfmu^\star,\bfSigma^\star$) instead of point estimates. In a similar way, considering \begin{equation*}
    p(\theta\vert Z) =\int p(\theta\vert Z,W) p\left(W\vert Z\right)dW
\end{equation*} step \ref{fcshomogen1c} yields a draw of $\bftheta^{(\nbiter)}$ from $p(\bftheta\vert Z)$ using the previous draw of $W$ given $Z$.
Regarding Step 2, it yields a draw of $W^{(\nbiter)}$ in its conditional distribution given $\bftheta^{(\nbiter)}$. After both steps, $\left(\bftheta,W\right)$ are well simulated under their joint distribution given $Z$ at iteration $\nbiter$. Next, knowing the categorical variable, sequential imputation of continuous variables can be performed and so on. Details about the algorithm are provided in Appendix \ref{FCS-homoalgo}.\\

The tuning parameters of this algorithm are the number of clusters $\Nbgroup$ and the number of iterations $\Nbiter$. The number of clusters can be estimated as
\begin{equation}\label{nbdegroup}
    argmin_{\Nbgroup\in \{1,...,\Nbgroup_{max}\}}\ T_\Nbgroup
\end{equation}
where $\Nbgroup_{max}$ is a bound for the number of clusters and $T_\Nbgroup$ is the instability obtained by after MI (Equation \ref{B3}) for a clustering in $\Nbgroup$ clusters. The number of iterations needs to be sufficiently large for ensuring convergence to a stationary distribution. Monitoring convergence is usually done graphically by investigating values of statistics (like average, variance, correlation, ...) during the iterative process. 

\subsection{Accounting for congeniality \label{fcs-hetero}}
To ensure congeniality, the imputation model needs to be tuned to account for the cluster analysis model. When, the analyst aims to fit a heteroscedastic gaussian mixture model, the previous imputation method is theoretically irrelevant. However, it can be easily modified to accounting for heteroscedasticity. For achieving this goal, imputation of continuous variables 
are replaced by fitting a heteroscedastic regression model (see  \cite{Audigier18} for a recent review). For generating the cluster variable, step \ref{fcshomogen1a} 
is accommodated by fitting a heteroscedastic gaussian mixture model, while linear discriminant functions in Steps \ref{fcshomogen1b} and \ref{fcshomogen2} 
are replaced by quadratic discriminant functions. Details are provided in Appendix \ref{FCS-heteroalgo}. As a consequence, the imputation model will be more complex, allowing a better fit of the model, but requiring more observations. Congenial properties of FCS MI methods (homoscedatic or heteroscedastic) are summarized in Table \ref{tab:summaryjm}. 

Beyond the heteroscedasticity, the previous FCS method can be tuned to account for other data structures. For instance, the imputation of continuous variables can be performed using other imputation methods including non-parametric methods like random forest \citep{Doove14} or semi-parametric like predictive mean-matching \citep{Morris14,Gaffert18}. 

\begin{table}[H]
\centering
\footnotesize
\caption{Summary of congenial properties for MI methods\label{tab:summaryjm}}
\begin{tabular}{|l|c|p{1.2cm}|p{1.2cm}|c|p{1.2cm}|p{1.2cm}|p{1.2cm}|p{1.2cm}|}
\hline     & \multicolumn{4}{c}{JM-GL / FCS-homo} & \multicolumn{4}{|c|}{ JM-DP / FCS-hetero}\\ \hline
Imputation model & \multicolumn{4}{c}{ homo}&\multicolumn{4}{|c|}{ hetero}\\ \hline     
Analysis model& \multicolumn{2}{c}{homo} &\multicolumn{2}{|c}{ hetero}& \multicolumn{2}{|c}{ homo}&\multicolumn{2}{|c|}{ hetero}\\ \hline
Truth&homo&hetero&homo&hetero&homo&hetero&homo&hetero\\ \hline
Congenial & yes & yes (but invalid) &no (but superefficient)&no (biased) &no (conservative)&no (biased)&yes (but biased)&yes\\ \hline
\end{tabular}
\end{table}

\section{Simulations\label{Section4}}
\subsection{Simulation design}
The previous FCS MI methods for clustering are assessed through a simulation study. According to previous comments, various configurations of the data are investigated. For achieving this goal, one of the configurations is considered as base-case, while the others vary by the separability of the clusters, the number of clusters, the cluster size, the proportions of each cluster as well as the covariance structure in each cluster. For each configuration, $\Nbsim=200$ data sets are generated and for each of them, various missing data patterns are considered. Then, MI methods are applied ($\Nbtab=20$) and clustering is performed using various algorithms. For each of them, partitions are pooled using non-negative matrix factorization \citep{Li07, Audigier20}. Finally, the accuracy of the pooled partitions is assessed by the adjusted rand index. The R code used for simulations is available on demand.
\subsubsection{Data generation\label{secdatagen}}
Data are generated according to a finite gaussian mixture model. The base-case configuration considers $\Nbgroup=3$ components, a number of individuals in each cluster $\Nbind_\nbgroup=250$ (for all $\nbgroup$ in $\{1,2,3\}$) and $\Nbvar=8$ variables. The expected mean of each component $\mu_\nbgroup$ is as follows:
\begin{eqnarray*}
\mu_1&=&(0, 0, 0, 0,  \Delta, \Delta, 0, \Delta^2)\\
\mu_2&=&(0, 0, 0, 0,  -\Delta, -\Delta, -\Delta, 0) \\
\mu_3&=&(0, 0, 0, 0,  -\Delta, \Delta, \Delta,  -\Delta^2)
\end{eqnarray*} with $\Delta=2$, while the covariance matrix $\Sigma_\nbgroup=\Sigma(\rho)={\scriptsize{\left(\begin{array}{cc}
    I_4 &  \text{\huge0}\\
    \text{\huge0} & \begin{array}{cccc}1&\rho&\rho&\rho\\
    \rho&1&\rho&\rho\\
    \rho&\rho&1&\rho\\
    \rho&\rho&\rho&1
    \end{array}
\end{array}\right)}}$ with $\rho=0.3$ for all $\nbgroup$ in $\{1,2,3\}$. 10 other configurations are investigated by varying the separability between clusters (model-II and model-III), the number of clusters (model-IV and model-V), the cluster size (model-VI and model-VII), the balance between clusters sizes (model-VIII and model-IX) and the heteroscedasticity (model-X and model-XI). Details are given in Table \ref{tab:config} in Appendix. 

For each generated data set, missing values are generated according to 9 mechanisms varying by the percentage of missing values $\tau$ (10\%, 25\%, 40\%) and the type of mechanisms (MCAR, MAR 1, MAR 2). Mechanisms of type MAR 1 are unrelated with the cluster structure, contrary to mechanisms of type MAR 2. More precisely, the distribution of the mechanisms are as follows:
$Prob(r_{\nbind\nbvar}=0)=\tau$ (MCAR mechanism) or $Prob(r_{\nbind\nbvar}=0)=\Phi(a_\tau+x_{\nbind 1})$ (MAR 1) or $Prob(r_{\nbind\nbvar}=0)=\Phi(a_\tau+x_{\nbind 8})$ (MAR 2) with $\Phi$ the cumulative distribution function of the standard normal distribution and $a_\tau$ a constant to control the percentage of missing values in expectation. Since the distribution of $X_1$ is the same in all clusters, the distribution of missing values is the same in all clusters for mechanisms of type MAR 1. On the contrary, since the distribution of $X_8$ varies according to the cluster some clusters have more missing values than others under mechanisms of type MAR 2. 

Note the MAR mechanisms (MAR 1 or MAR 2) are more prone to generate missing values on the same individuals, which corresponds to usual cases in practice. Such missing data patterns are more difficult for clustering since if all variables related to the cluster structure are missing, then individuals cannot be suitably clustered (and imputed).


\subsubsection{Imputation methods\label{misim}}
On each incomplete data set, various MI methods are applied :
\begin{itemize}
    \item FCS-homo: FCS MI as proposed in Section \ref{fcs-homo} using $\Nbiter=200$ and $\Nbgroup$ set to the true number of clusters
    \item FCS-hetero: MI by the FCS imputation proposed in Section \ref{fcs-hetero} using $\Nbiter=200$ and $\Nbgroup$ set to the true number of clusters
    \item FCS-norm: FCS MI under the linear gaussian model \citep{VB18} using $\Nbiter=20$ iterations. This MI method ignores the latent class structure.
    \item JM-GL: JM MI under the general location model \citep{Schafer97, mixpackage}. $\Nbgroup$ is set to the true number of clusters. 100 iterations are used for the burn-in step and 20 between saved imputations.
    \item JM-DP: JM MI based on the Dirichlet process mixture of products of multivariate normal distributions \citep{Kim14, Kimpackage}. Hyperparameters are set as follows $h=1$, $\mu_0=0$, $df=\Nbvar+1$, $a_0=0.25$, $b_0=0.25$, $a_\alpha=0.25$, $b_\alpha=0.25$. The number of components is bounded by 5 so that the range for the number of components is centered around the true number for the base-case configuration. 500 iterations are used for the burn-in step and 100 between saved imputations.
    \item JM-norm: MI by JM under the normality assumption \citep{Schafer97, Schafernormpkg}. 500 iterations are used for the burn-in step and 100 between saved imputations. This MI method ignores the latent class structure.
\end{itemize}

\noindent As benchmark, the following other methods are investigated:
\begin{itemize}
    \item Full: cluster analysis from full data
    \item Full-bagging: cluster analysis from full data using a bagging procedure \citep{Dudoit03}. Such a method is known to improve partition accuracy compared to clustering without bagging.
    \item SI: results for single imputation ($\Nbtab=1$). Since it avoids pooling, results from single imputation is only related to the fit of the imputation model, while MI additionally depends on the independence between partitions.
\end{itemize}

\subsubsection{Cluster analyses}
On each imputed data set, various cluster analysis methods are considered
\begin{itemize}
    \item mixture: clustering by gaussian mixture model. The covariance matrix is parametrized according to the covariance structure used for data generation.
    \item k-means: k-means clustering with euclidean distance with standardization.
    \item pam: partitioning around medoids with euclidean distance and standardization.
    \item hc: hierarchical clustering with euclidean distance using the Ward's method. Variables are standardized.
\end{itemize}
For each one, the number of clusters is set to the number used for data generation.

\subsubsection{Criteria}
For a given data set, a given imputation method and a given cluster analysis method, the adjusted rand index between the estimated partition and the true one is used for assessing the clustering accuracy.

\subsection{Results}

\subsubsection{Influence of the imputation model}
\begin{figure}
\begin{subfigure}[b]{.5\linewidth}
    \centering
    \includegraphics[scale=.35]{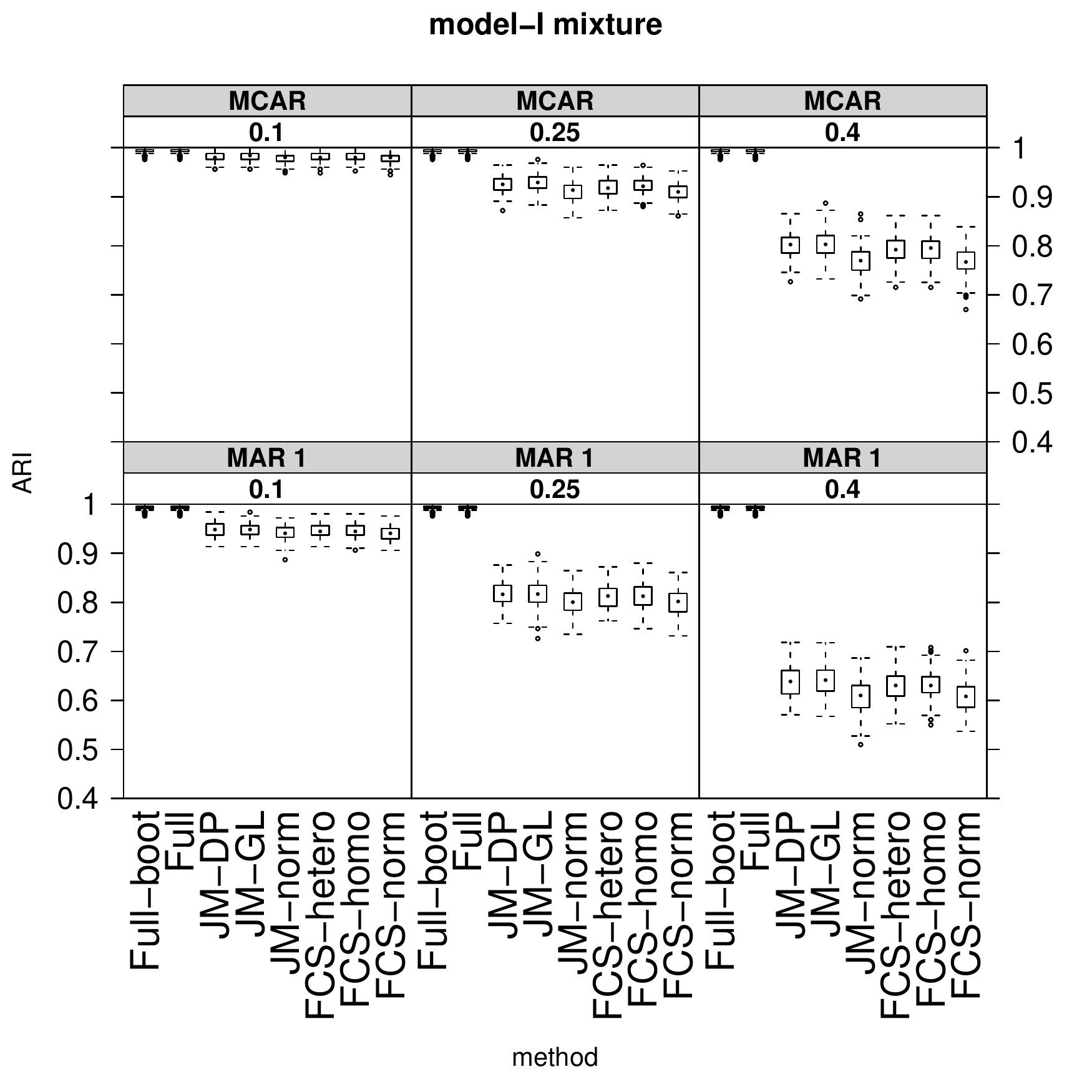}
    \caption{ARI distribution over the 200 data sets according to the type of missing data mechanism (MCAR or MAR 1) as well as the proportion of missing values (10\%, 25\%, 40\%). As benchmark, ARI for full data case are also reported (Full-boot, Full). \label{fig:refmi}}
    \end{subfigure}~~~\begin{subfigure}[b]{.5\linewidth}
    \centering
    \includegraphics[scale=.35]{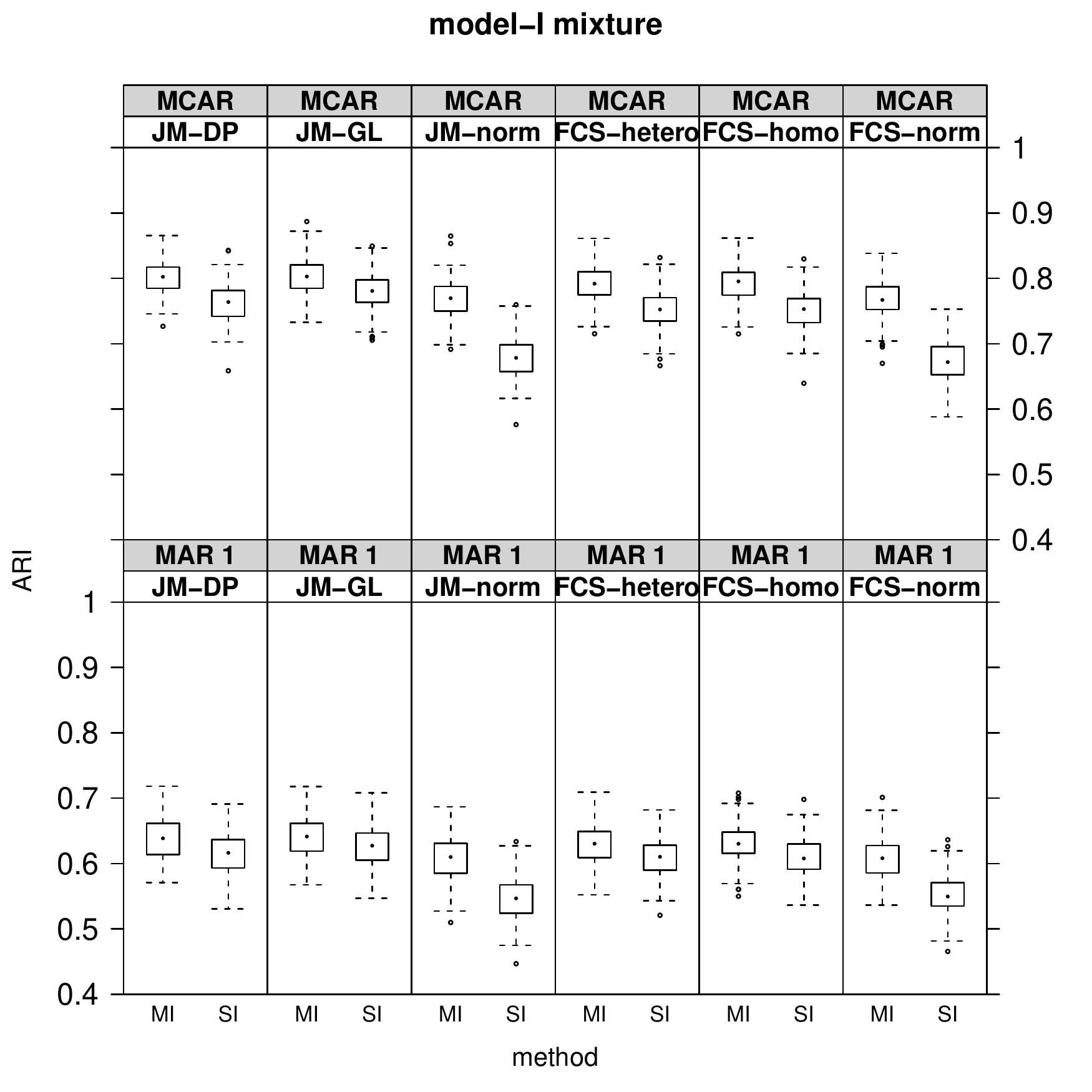}
    \caption{ARI distribution over the 200 data sets according to the type of missing data mechanism (MCAR or MAR 1) for 40\% of missing values. \label{fig:refsi}\\ \\}
    \end{subfigure}
    \caption{Base-case configuration: comparison between imputation models (\ref{fig:refmi}) and comparison between MI and SI (\ref{fig:refsi}) for various imputation methods (JM-DP, JM-GL, JM-norm, FCS-hetero, FCS-homo, FCS-norm). Cluster analysis is performed using gaussian mixture models.}
    \label{fig:ref}
\end{figure}

ARI distributions over the $\Nbsim$ generated data sets are reported in the following figures as boxplots. See Tables \ref{tab:I}-\ref{tab:XI} in Appendix for medians and interquartile ranges. Since results under MAR 1 and MAR 2 mechanisms are generally close, in almost all configurations only results under MAR 1 are presented. In the same spirit, because differences between methods are very small with 10\% of missing values, the associated ARI distributions are generally not reported.
\paragraph{Base-case configuration}
Figure \ref{fig:refmi} reports the ARI distribution under the several missing data mechanisms. For all MI methods, the ARI decreases when the proportion of missing values increases. Furthermore, ARI are globally smaller under a MAR mechanism than under a MCAR mechanism. This behavior was expected since the MAR mechanism tends to concentrate missing values on a smaller number of individuals, which become difficult to cluster. The differences between MI methods are more pronounced when the proportion of missing values is large. In such a case, the MI methods accounting for the cluster structure (JM-DP, JM-GL, FCS-homo and FCS-hetero) outperform the others (FCS-norm and JM-norm). Among them, JM MI methods provide slightly higher ARI.


Furthermore, homoscedastic MI methods (JM-GL, FCS-homo) provide similar ARI than heteroscedastic methods (JM-DP, FCS-hetero), but we can note that interquartile ranges are generally slightly higher for heterosdastic methods (see Table \ref{tab:I} in Appendix). It was expected since, in this base-case configuration, data support constant variance for all clusters, meaning heteroscedatic imputation methods tend to be more conservative.

Comparison between MI and SI is presented in Figure \ref{fig:refsi} (only the case with 40\% of missing values is shown). It highlights that compared to SI, MI substantially improves ARI for all imputation methods.

\begin{figure}
\begin{subfigure}[b]{.5\linewidth}
    \centering
    \includegraphics[scale=.35]{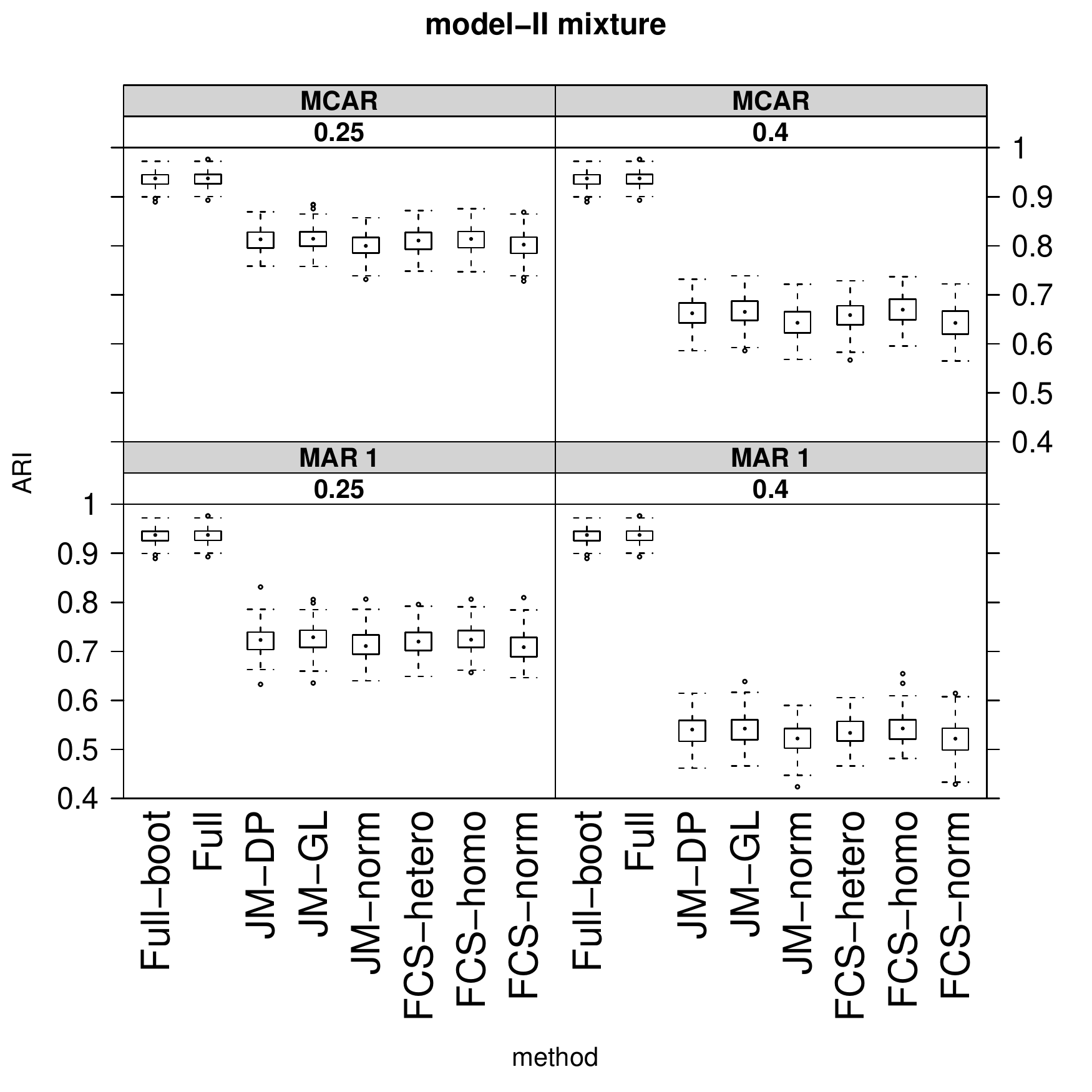}
    \caption{model-II (lower separability)}
     \end{subfigure}~~~\begin{subfigure}[b]{.5\linewidth}
    \centering
    \includegraphics[scale=.35]{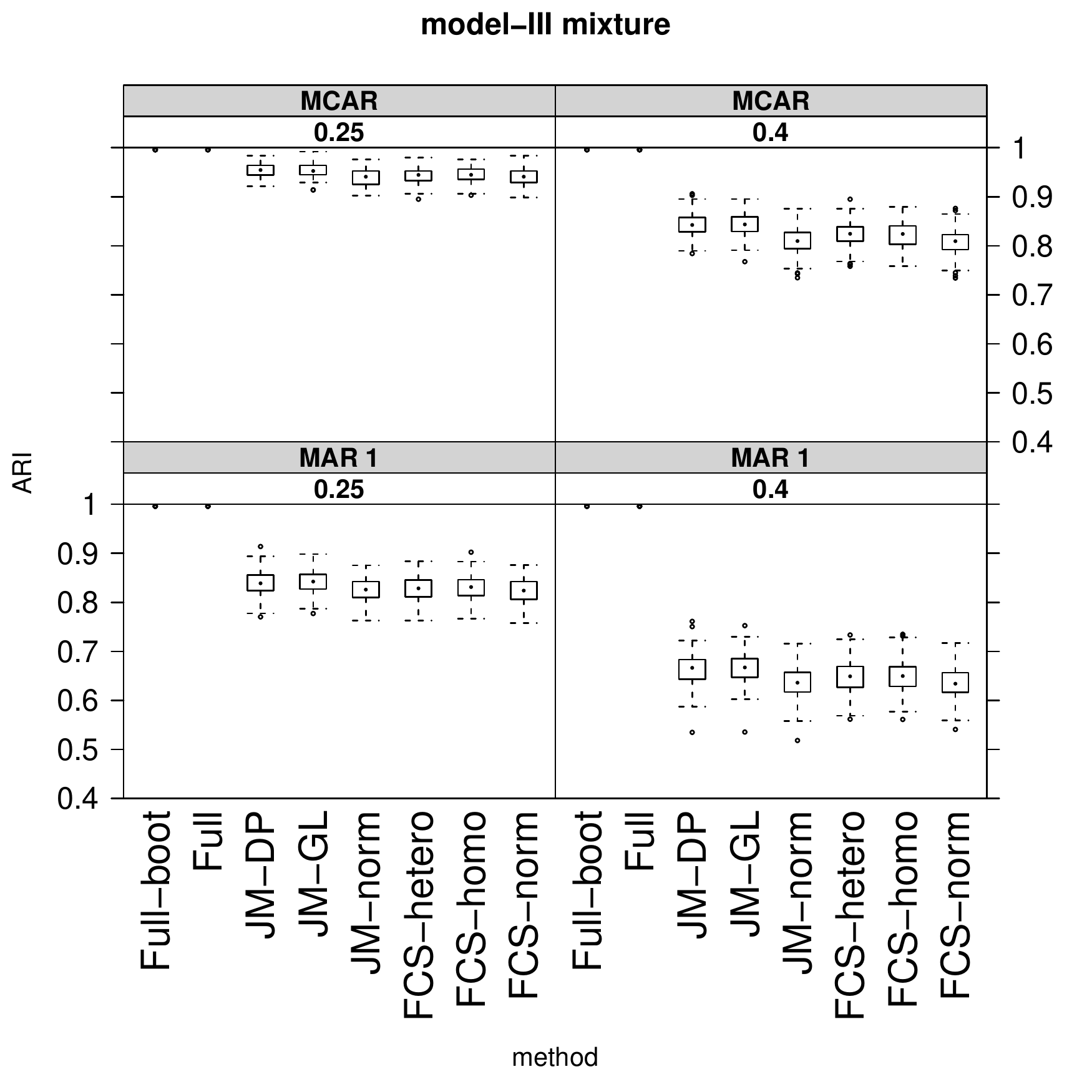}
    \caption{model-III (higher separability)}
     \end{subfigure}
    \caption{ARI distribution over the 200 data sets according to the type of missing data mechanism (MCAR or MAR 1) and the proportion of missing values (25\%, 40\%) for various MI methods (JM-DP, JM-GL, JM-norm, FCS-hetero, FCS-homo, FCS-norm). As benchmark, ARI for full data case are also reported (Full-boot, Full). Cluster analysis is performed using gaussian mixture models.}
    \label{fig:sep}
\end{figure}

\paragraph{Separability}
Figure \ref{fig:sep} presents ARI when clusters are less separated (model-II) or more separated (model-III) compared to the base-case configuration. ARI are globally smaller when clusters are less separated, as observed for full data cases.

\begin{figure}
    \begin{subfigure}[b]{.5\linewidth}
    \centering
    \includegraphics[scale=.35]{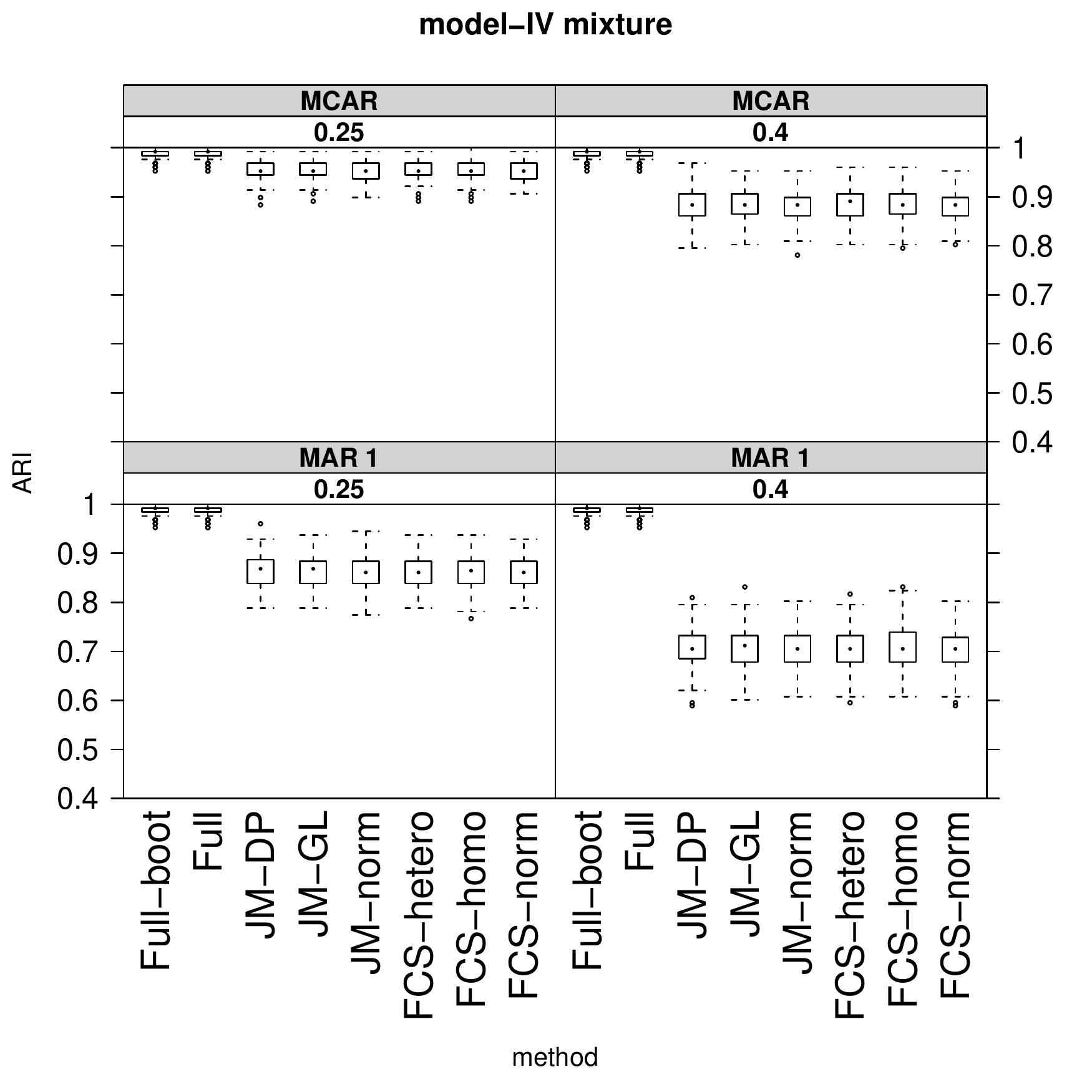}
    \caption{model-IV (two clusters)\label{fig:nbgp2}}
     \end{subfigure}~~~\begin{subfigure}[b]{.5\linewidth}
    \centering
    \includegraphics[scale=.35]{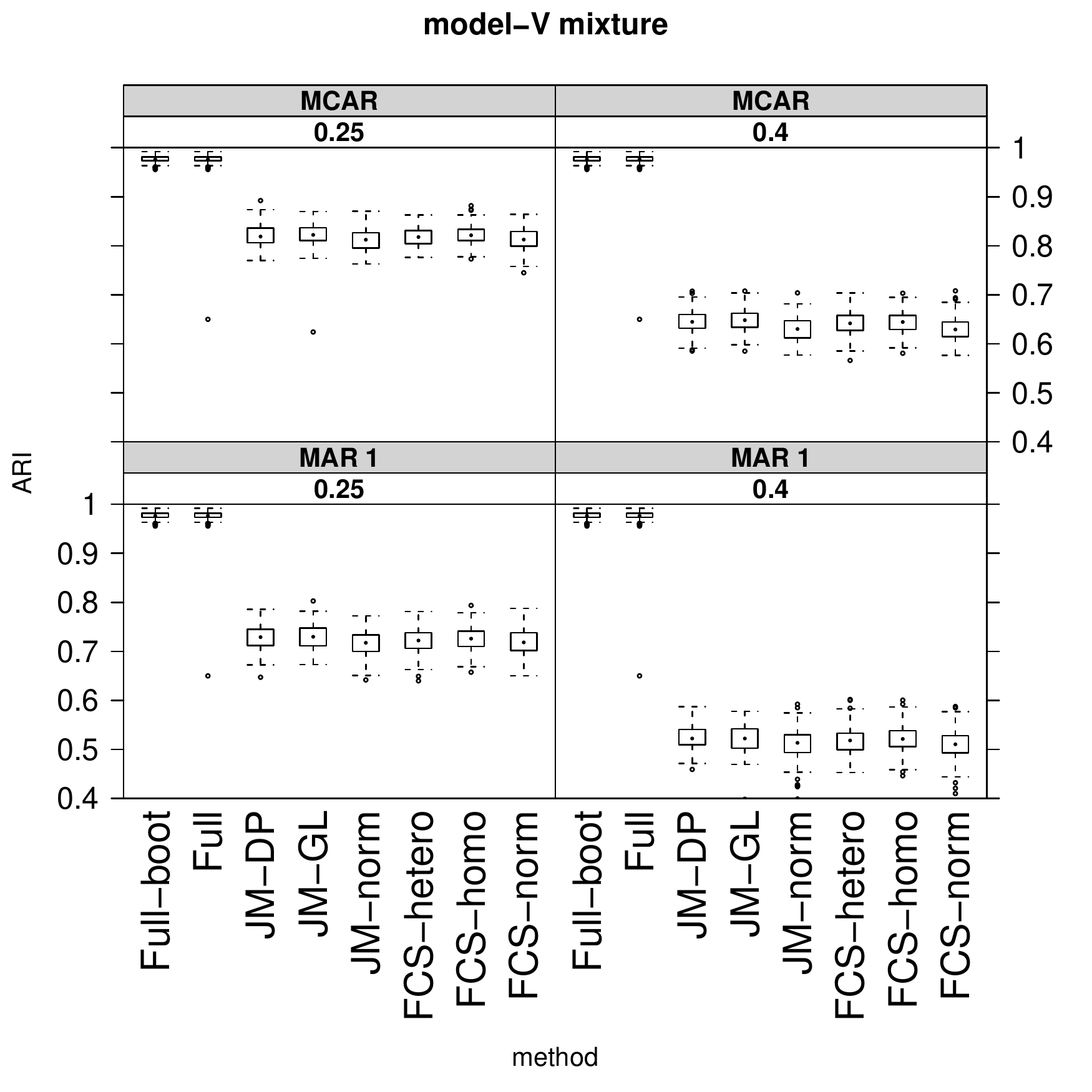}
    \caption{model-V (four clusters)\label{fig:nbgp4}}
     \end{subfigure}
    \caption{ARI distribution over the 200 data sets according to the type of missing data mechanism (MCAR or MAR 1) and the proportion of missing values (25\%, 40\%) for various MI methods (JM-DP, JM-GL, JM-norm, FCS-hetero, FCS-homo, FCS-norm). As benchmark, ARI for full data case are also reported (Full-boot, Full). Cluster analysis is performed using gaussian mixture models.}
    \label{fig:nbgp}
\end{figure}

\paragraph{Number of clusters}
Figure \ref{fig:nbgp} reports results for a smaller number of clusters (model-IV on the left) and a higher number (model-V on the right). It should be noticed that ARI on model-IV and model-V cannot be directly compared since the ARI is sensitive to the number of clusters.  When they are only two clusters (model-IV) all MI methods perform similarly, even if the cluster structure is not considered, as in JM-norm and FCS-norm. The reason is that the coordinates of both centers of gravity are linearly dependent (while this is not true for $\Nbgroup>2$). In such a case, imputation by normal distribution is a sufficient approximation for clustering.

\begin{figure}
    \begin{subfigure}[b]{.5\linewidth}
    \centering
    \includegraphics[scale=.35]{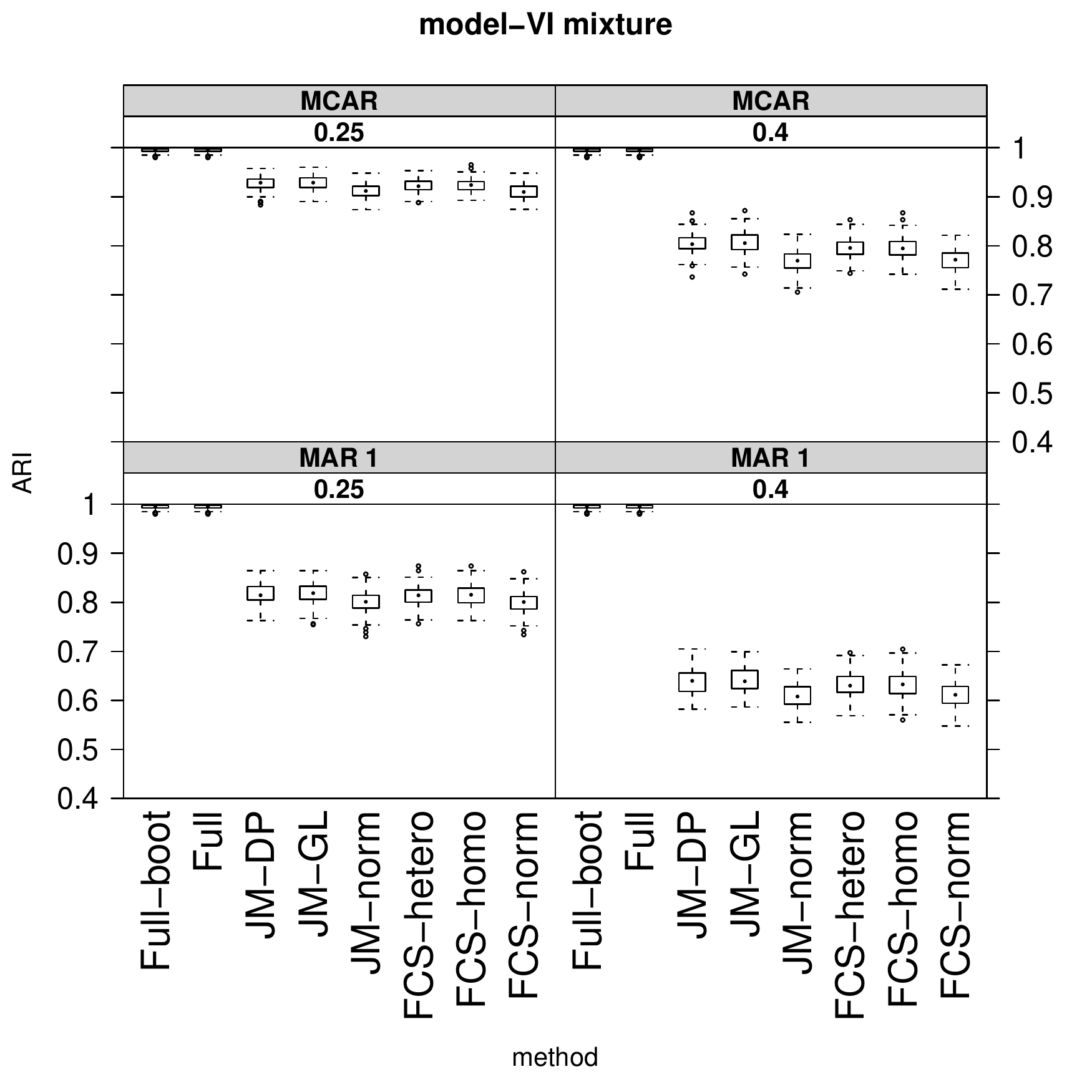}
    \caption{model-VI (400 individuals per cluster)}
     \end{subfigure}~~~\begin{subfigure}[b]{.5\linewidth}
    \centering
    \includegraphics[scale=.35]{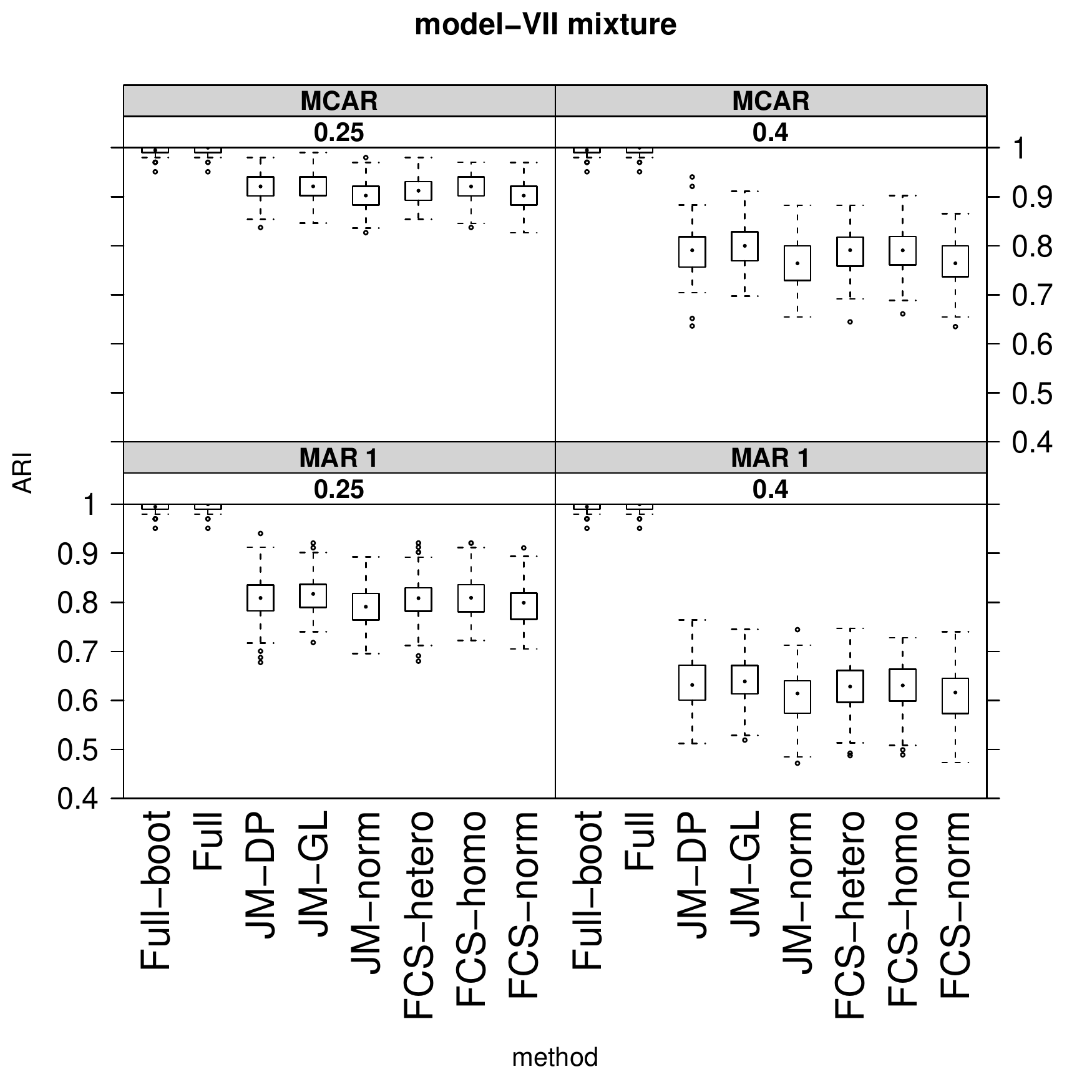}
    \caption{model-VII (100 individuals per cluster)}
     \end{subfigure}
    \caption{ARI distribution over the 200 data sets according to the type of missing data mechanism (MCAR or MAR 1) and the proportion of missing values (25\%, 40\%) for various MI methods (JM-DP, JM-GL, JM-norm, FCS-hetero, FCS-homo, FCS-norm). As benchmark, ARI for full data case are also reported (Full-boot, Full). Cluster analysis is performed using gaussian mixture models.}
    \label{fig:nbind}
\end{figure}

Note that JM-GL often fails to converge on model-V, consequently the corresponding boxplot concerns a smaller number of data sets (see Table \ref{tab:V} in Appendix for details).
\paragraph{Number of individuals}
Figure \ref{fig:nbind} gathers results when the number of individuals is higher (model-VI on the left) or smaller (model-VII on the right) than the number considered in the base-case configuration (400 and 100 respectively). For a given MI method, the ARI is robust to the number of individuals in average, but the interquartile range is two times larger on model-VII than on model-VI.

\begin{figure}
    \begin{subfigure}[b]{.5\linewidth}
    \centering
    \includegraphics[scale=.35]{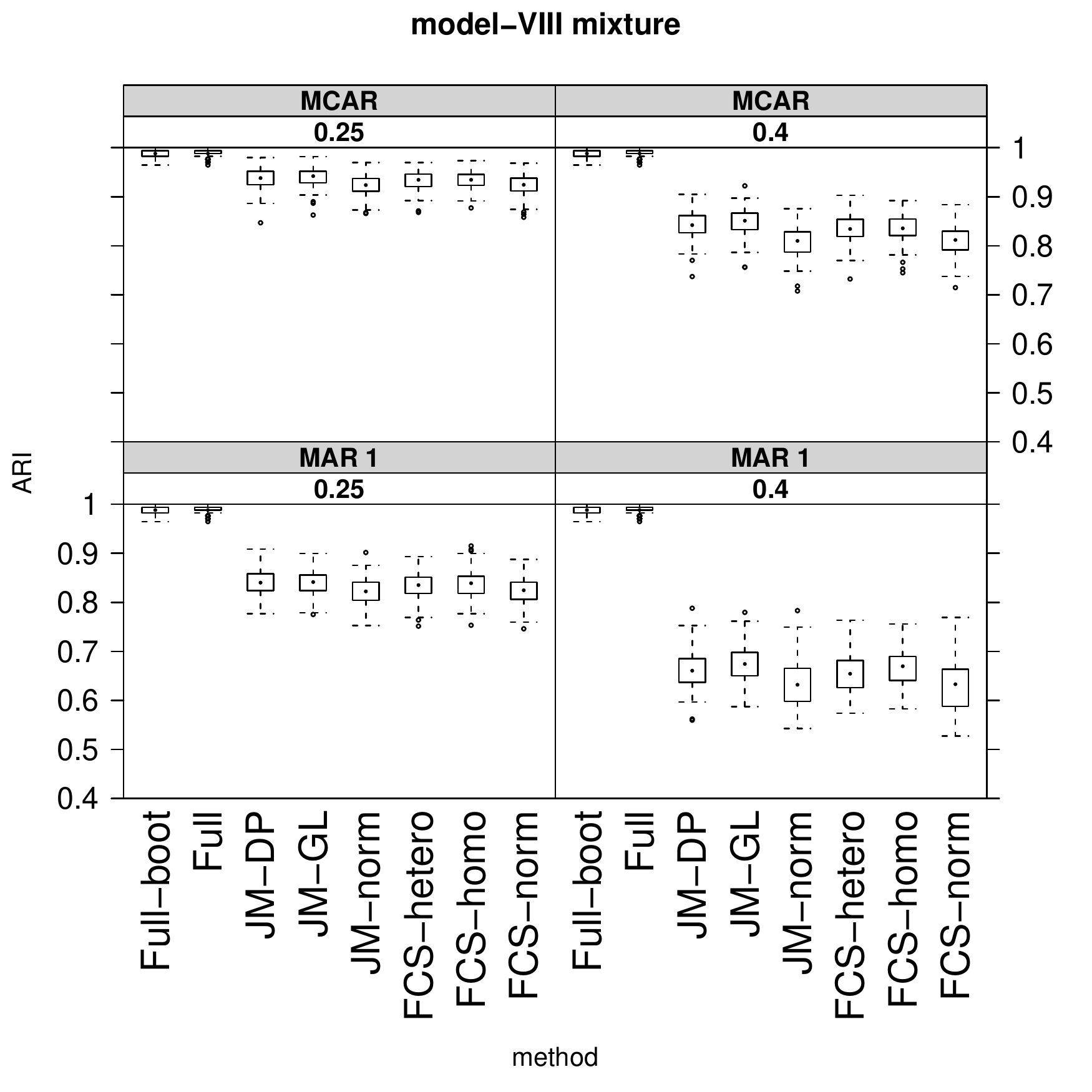}
    \caption{model-VIII (100 individuals in one cluster)}
     \end{subfigure}~~~ \begin{subfigure}[b]{.5\linewidth}
    \centering
    \includegraphics[scale=.35]{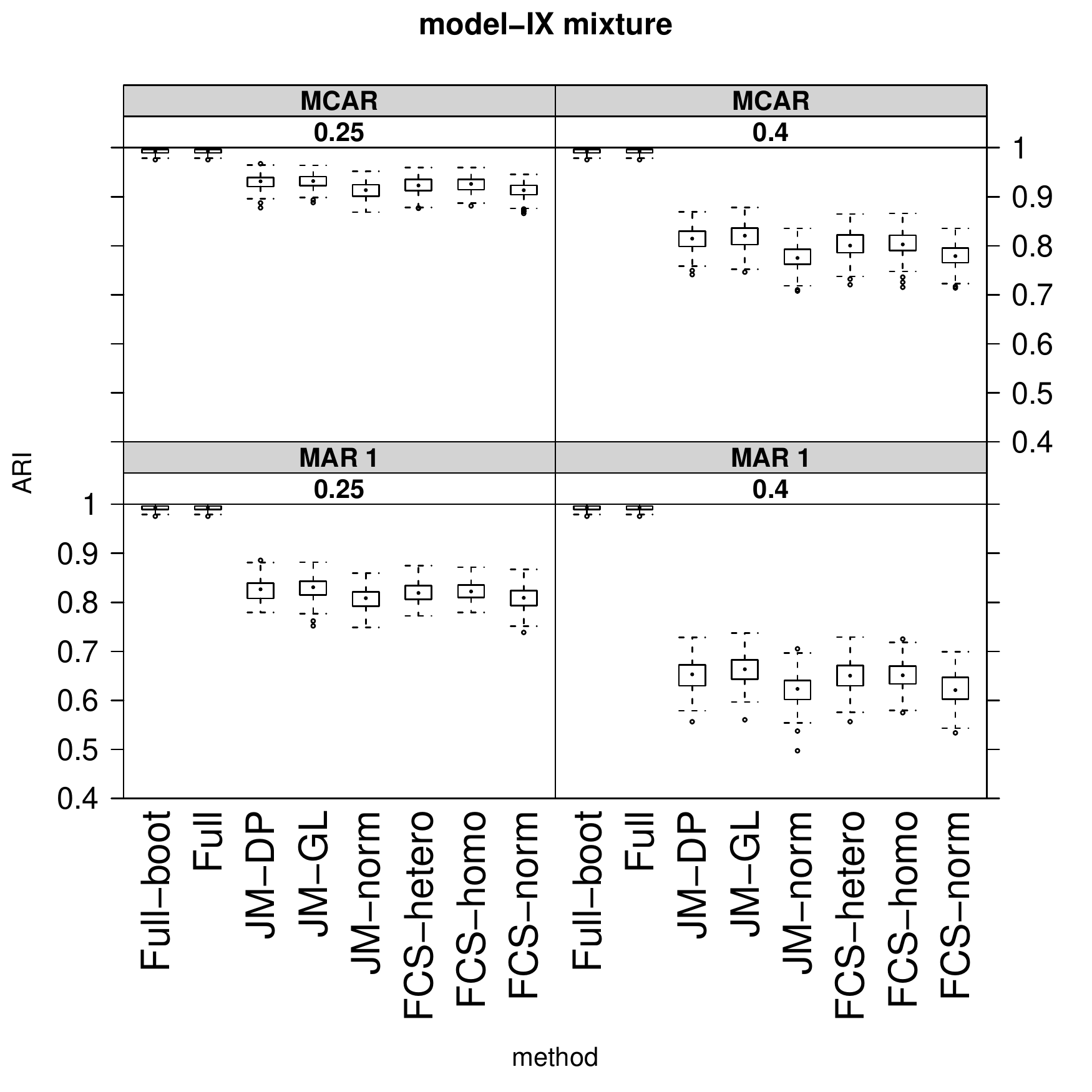}
    \caption{model-IX (400 individuals in one cluster)\\}
     \end{subfigure}
    \caption{ARI distribution over the 200 data sets according to the type of missing data mechanism (MCAR or MAR 1) and the proportion of missing values (25\%, 40\%) for various MI methods (JM-DP, JM-GL, JM-norm, FCS-hetero, FCS-homo, FCS-norm). As benchmark, ARI for full data case are also reported (Full-boot, Full). Cluster analysis is performed using gaussian mixture models.}
    \label{fig:balance}
\end{figure}

\paragraph{Unbalanced clusters}
Figure \ref{fig:balance} highlights ARI when one cluster is larger than the others (model-VIII on the left) or smaller (model-IX on the right). Similarly to Figure \ref{fig:nbind}, differences in terms of interquartile range can be noted, but these differences seem mainly related to the higher number of individuals on model-IX than on model-VIII.

\begin{figure}
         \begin{subfigure}[b]{.5\linewidth}
    \centering
        \includegraphics[trim={.2cm 0cm .5cm 0cm}, clip,scale=.35]{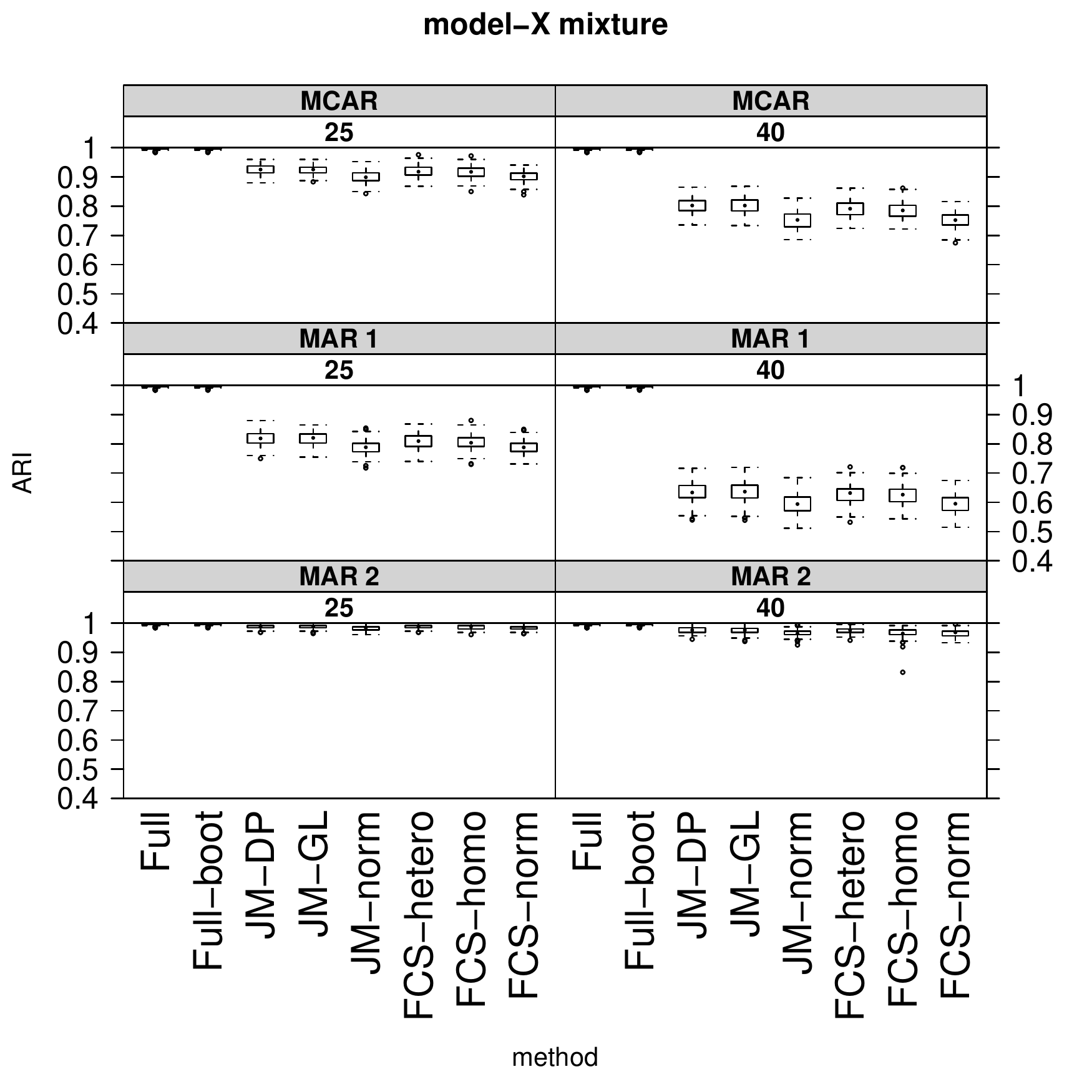}
    \caption{model-X}
     \end{subfigure}~~~\begin{subfigure}[b]{.5\linewidth}
    \centering
        \includegraphics[trim={.2cm 0cm .5cm 0cm},clip,scale=.35]{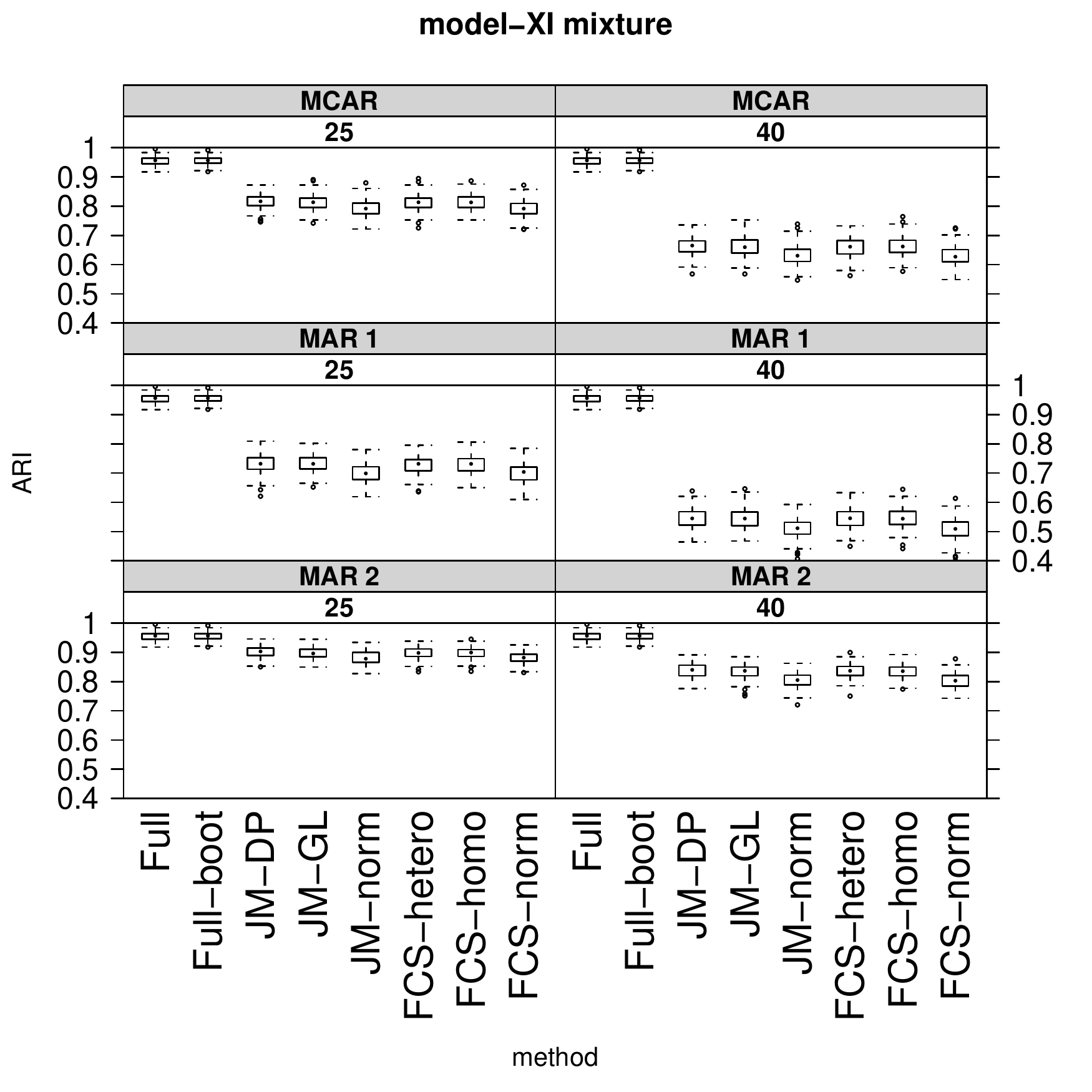}
    \caption{model-XI}
     \end{subfigure}
\caption{ARI distribution over the 200 data sets according to the type of missing data mechanism (MCAR, MAR 1 or MAR 2) and the proportion of missing values (25\%, 40\%) for various MI methods (JM-DP, JM-GL, JM-norm, FCS-hetero, FCS-homo, FCS-norm). As benchmark, ARI for full data case are also reported (Full-boot, Full). Cluster analysis is performed using gaussian mixture models.}
    \label{fig:hetero}
\end{figure}

\begin{figure}
         \begin{subfigure}[b]{.5\linewidth}
    \centering
        \includegraphics[trim={.2cm 0cm .5cm 0cm}, clip,scale=.35]{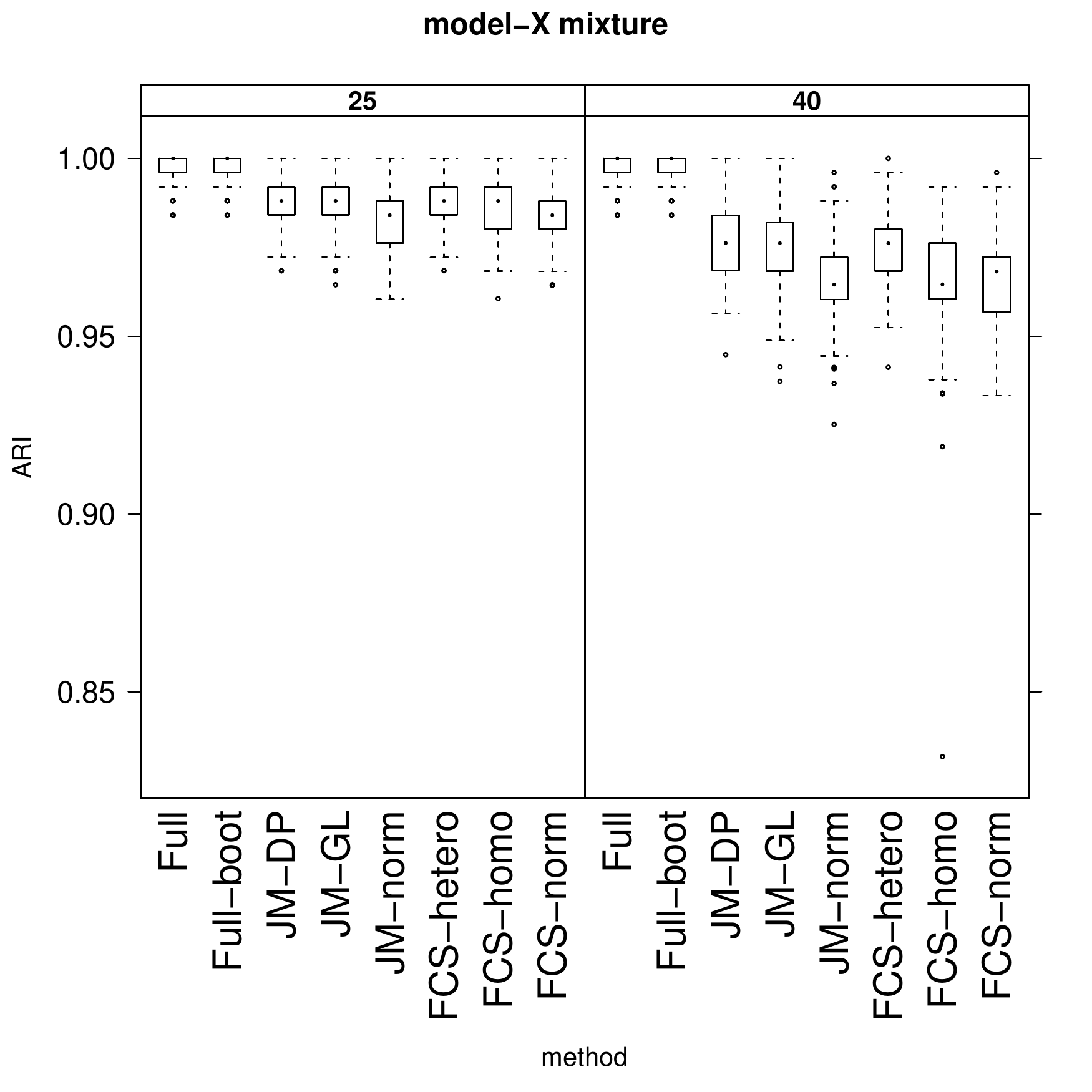}
    \caption{model-X}
     \end{subfigure}~~~\begin{subfigure}[b]{.5\linewidth}
    \centering
        \includegraphics[trim={.2cm 0cm .5cm 0cm},clip,scale=.35]{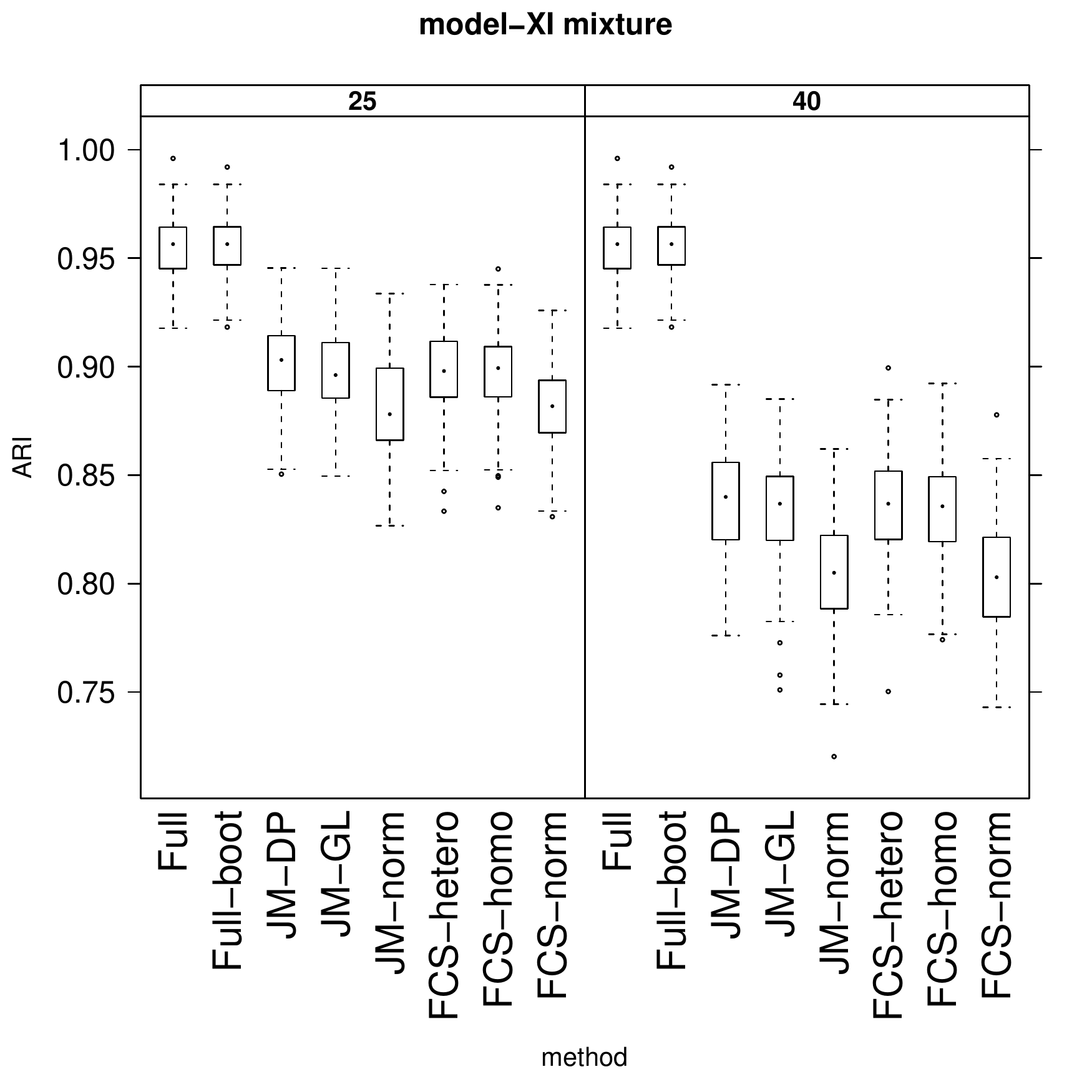}
    \caption{model-XI}
     \end{subfigure}
\caption{ARI distribution over the 200 data sets according to the proportion of missing values (25\%, 40\%) for various MI methods (JM-DP, JM-GL, JM-norm, FCS-hetero, FCS-homo, FCS-norm). Missing data are generated under the MAR 2 mechanism. As benchmark, ARI for full data case are also reported (Full-boot, Full). Cluster analysis is performed using gaussian mixture models. }
    \label{fig:heterozoom}
\end{figure}

\paragraph{Heteroscedasticity}
Figures \ref{fig:hetero} and \ref{fig:heterozoom} report simulations when data are generated according to heteroscedastic mixture models (Figure \ref{fig:heterozoom} focus on the MAR 2 mechanism by using a specific scale). Two cases are considered: model-X (on the left) where the cluster separation is the same as the base-case configuration and model-XI (on the right) where clusters are less separated (as in model-II). 

In both cases, homoscedastic MI methods (JM-GL and FCS-homo) are uncongenial with the analysis model, but they perform as well as heteroscedastic MI methods (JM-DP and FCS-hetero) when the missing data pattern is not related to the cluster structure (MCAR and MAR 1) mechanism. On the contrary, when the missing data pattern is strongly related to the cluster structure (MAR 2), heteroscedastic methods provide partitions with slightly highest ARI (see Figure \ref{fig:heterozoom}). 

We can note that ARI under MAR 2 mechanisms are higher than those observed under MAR 1. This is directly related to the full variable which is used to define the mechanism: under MAR 1, this variable is unrelated to the cluster structure, while under MAR 2 it strongly structure the data, meaning a full variable is available.

\subsubsection{Influence of the cluster analysis method}
\begin{figure}
    \begin{subfigure}[b]{.5\linewidth}
    \centering
    \includegraphics[trim={.2cm 0cm .5cm 0cm},clip,scale=.4]{model-I_mixture_soft.pdf}
    \caption{mixture}
    \end{subfigure}~~~\begin{subfigure}[b]{.5\linewidth}
    \centering
    \includegraphics[trim={.2cm 0cm .5cm 0cm},clip,scale=.4]{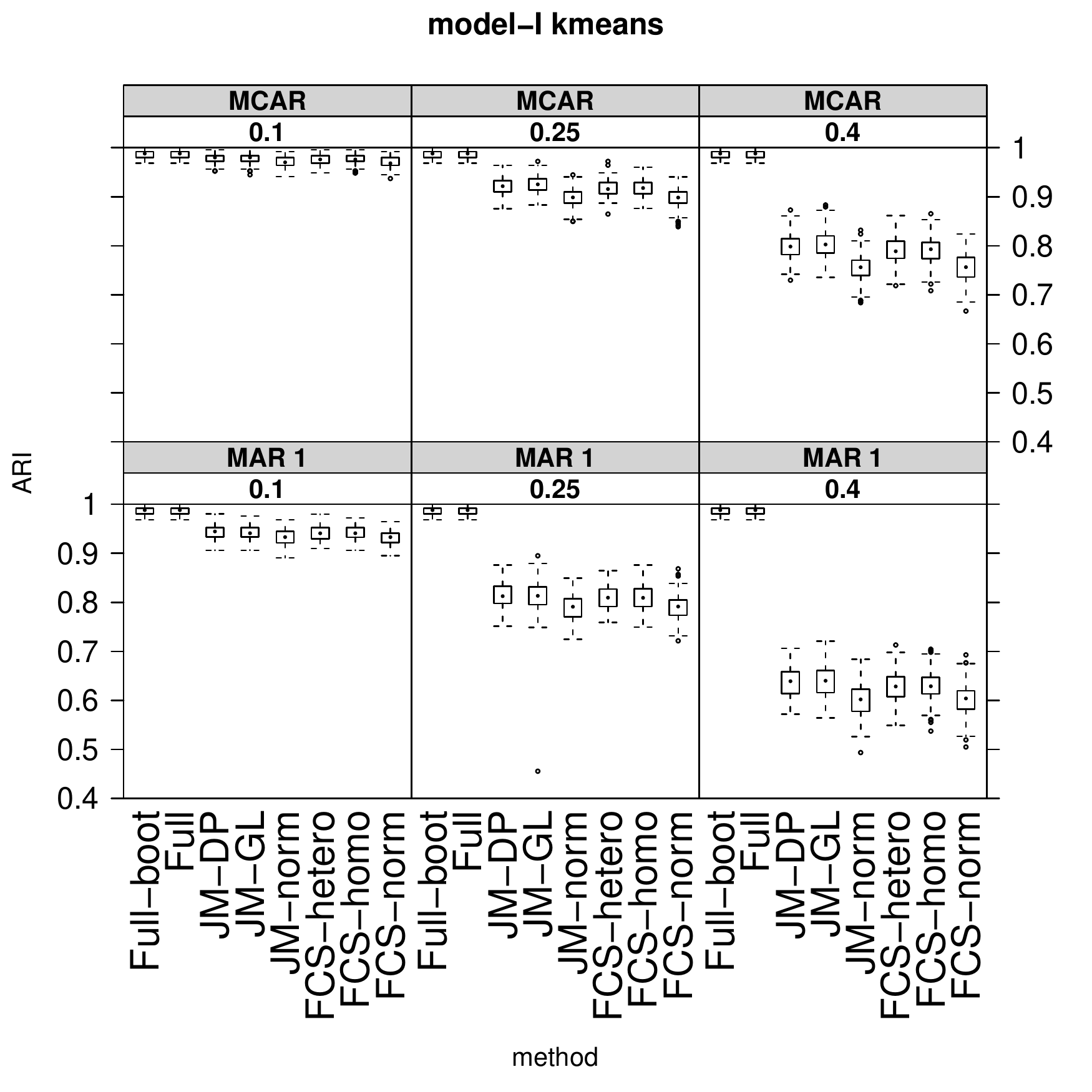}
    \caption{kmeans}
    \end{subfigure}
    \begin{subfigure}[b]{.5\linewidth}
    \centering
    \includegraphics[trim={.2cm 0cm .5cm 0cm},clip,scale=.4]{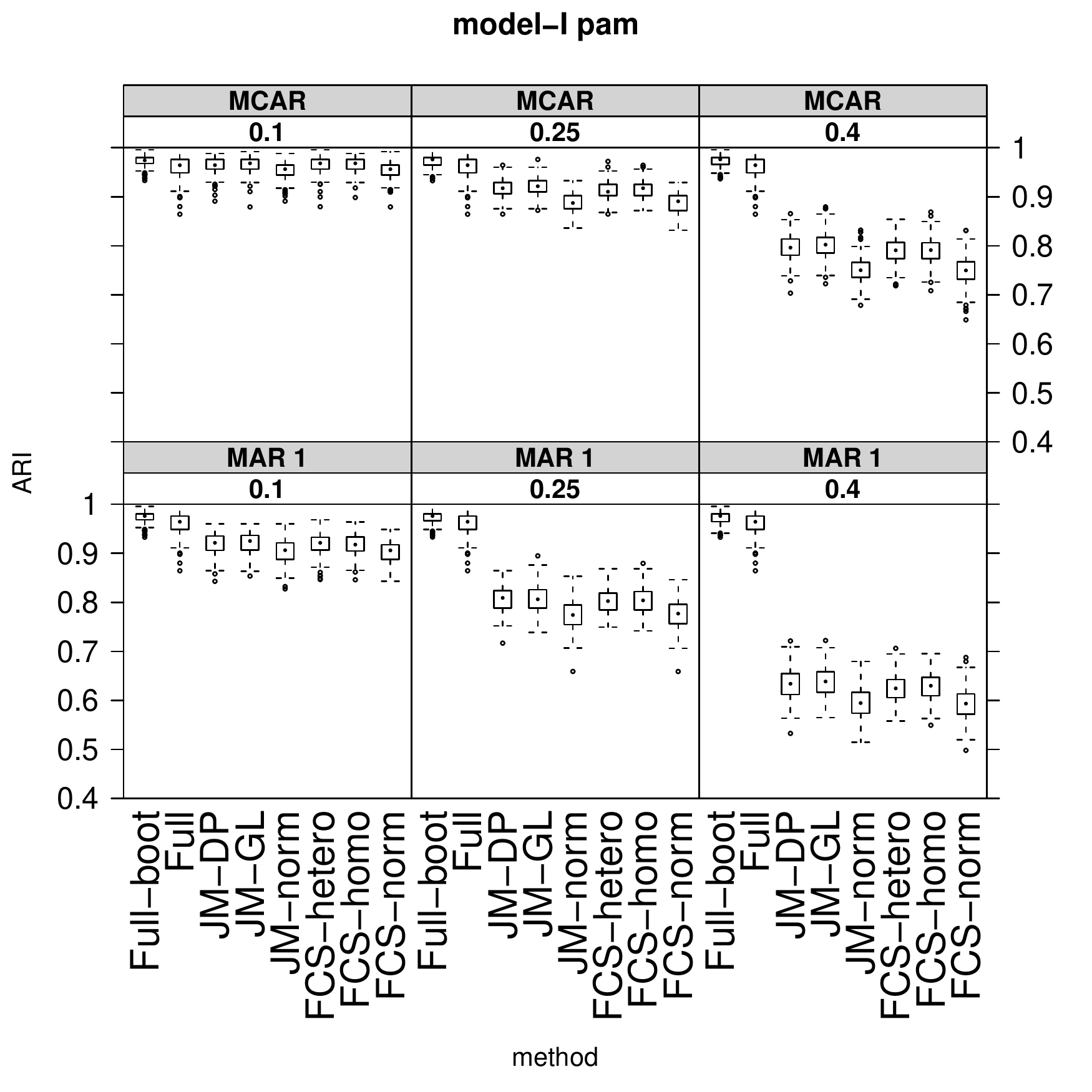}
    \caption{pam}
    \end{subfigure}~~~\begin{subfigure}[b]{.5\linewidth}
    \centering
    \includegraphics[trim={.2cm 0cm .5cm 0cm},clip,scale=.4]{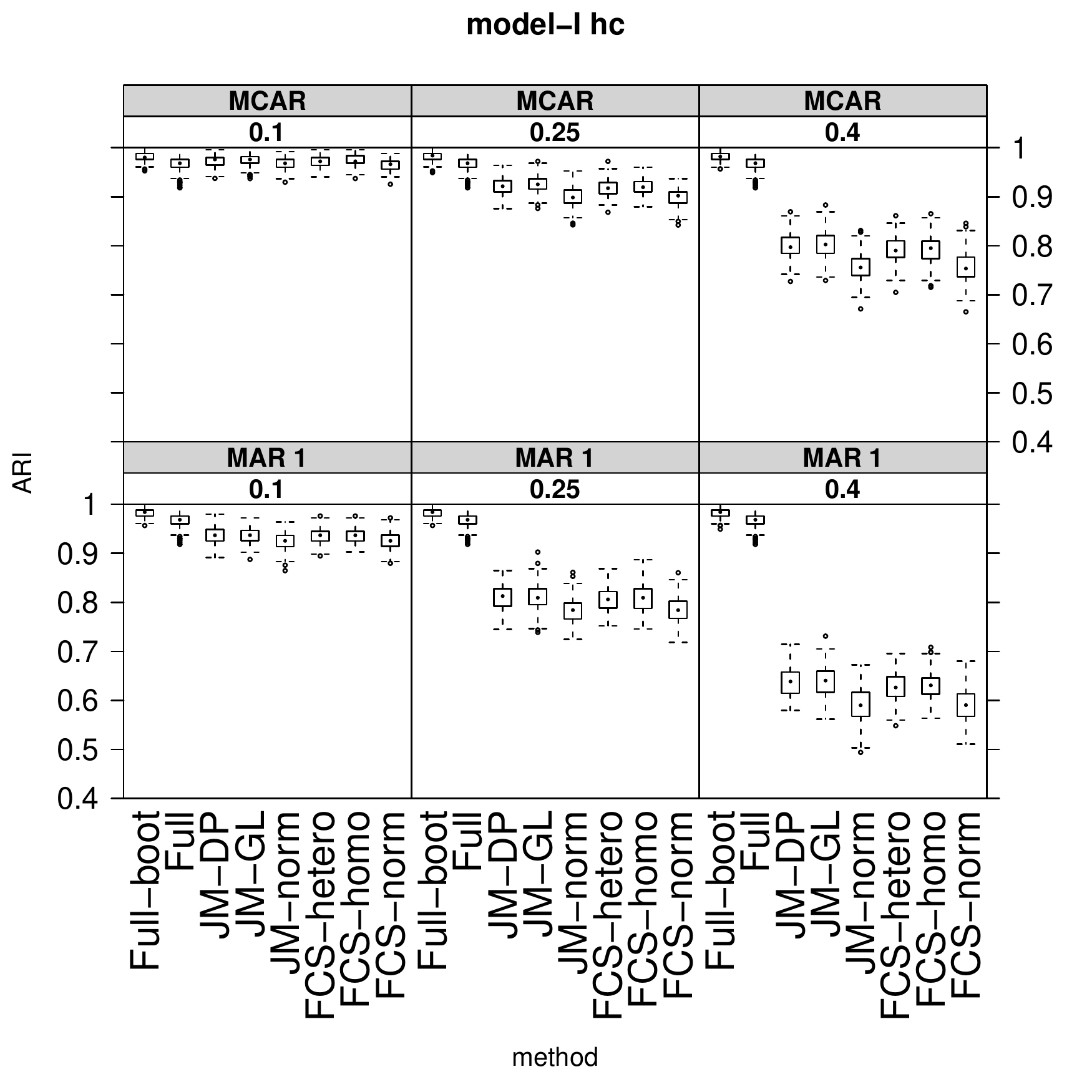}
    \caption{hierarchical clustering}
    \end{subfigure}
    \caption{Influence of the cluster analysis method for base-case configuration: ARI distribution over the 200 data sets according to the type of missing data mechanism (MCAR or MAR 1) and the proportion of missing values (10\%, 25\%, 40\%) for various MI methods (JM-DP, JM-GL, JM-norm, FCS-hetero, FCS-homo, FCS-norm). As benchmark, ARI for full data case are also reported (Full-boot, Full).}
    \label{fig:ana}
\end{figure}

Figure \ref{fig:ana} reports clustering performances on the base-case configuration for various cluster analyses methods. As expected, ARI when using k-means clustering or partitioning around medoids are similar to ARI obtained from gaussian mixture models. Hierarchical clustering, even if it cannot be directly related to mixture model provides also similar ARI. Interestingly, with 10\% of missing values under the MCAR mechanism, MI methods accounting for the class structure outperform clustering on full data (without bagging) when pam or hierarchical clustering is used. This illustrates the pooling step of MI can largely offset a moderate data loss.

\subsubsection{Summary}
These simulations results highlight ignoring the underlying class structure in the imputation model increases bias compared to dedicated MI methods. On the contrary, when the proportion of missing values is small, MI by dedicated methods provides similar or even better ARI than those observed for the full data case; when the proportion of missing values is large, the reference partition cannot be identified so well, because the observed profiles of each individual are not sufficient. Comparing homoscedastic and heteroscedastic imputation models, using congenial imputation models improves ARI, but the loss to use an heteroscedastic model instead of a homoscedastic model remains small (and vice versa).
\subsection{Wine data set}
The wine data set \citep{wine} is considered to illustrate the flexibility of FCS MI methods compared to JM MI methods for clustering. Such a data set describes $\Nbind=178$ Italian wines by $\Nbvar=13$ chemical descriptors. These wines can be clustered in $\Nbgroup=3$ categories (corresponding to 3 cultivars). Since the data set is full, 40\% missing values are added according to a MCAR or a MAR mechanism (as described in Section \ref{secdatagen}). 100 missing data patterns are considered for each mechanism.

JM MI methods are applied using the same parameters as those described in Section \ref{misim}. For FCS MI methods, variable selection is performed from each incomplete data to define each conditional imputation model. The method used is fully described in \citet{Barhen20}, it consists in repeating imputation and variable selection on subset of variables drawn at random. On each subset, imputation is performed using JM-GL and variable selection using knockoff \citep{Barber15}. 
 The number of clusters is set to $\Nbgroup=3$ for the proposed FCS MI methods. In practice, it could be estimated from incomplete values according to Equation \ref{nbdegroup}.
The number of iterations is tuned to $\Nbiter=200$.

\begin{figure}[H]
   \centering
    \includegraphics[scale=.5]{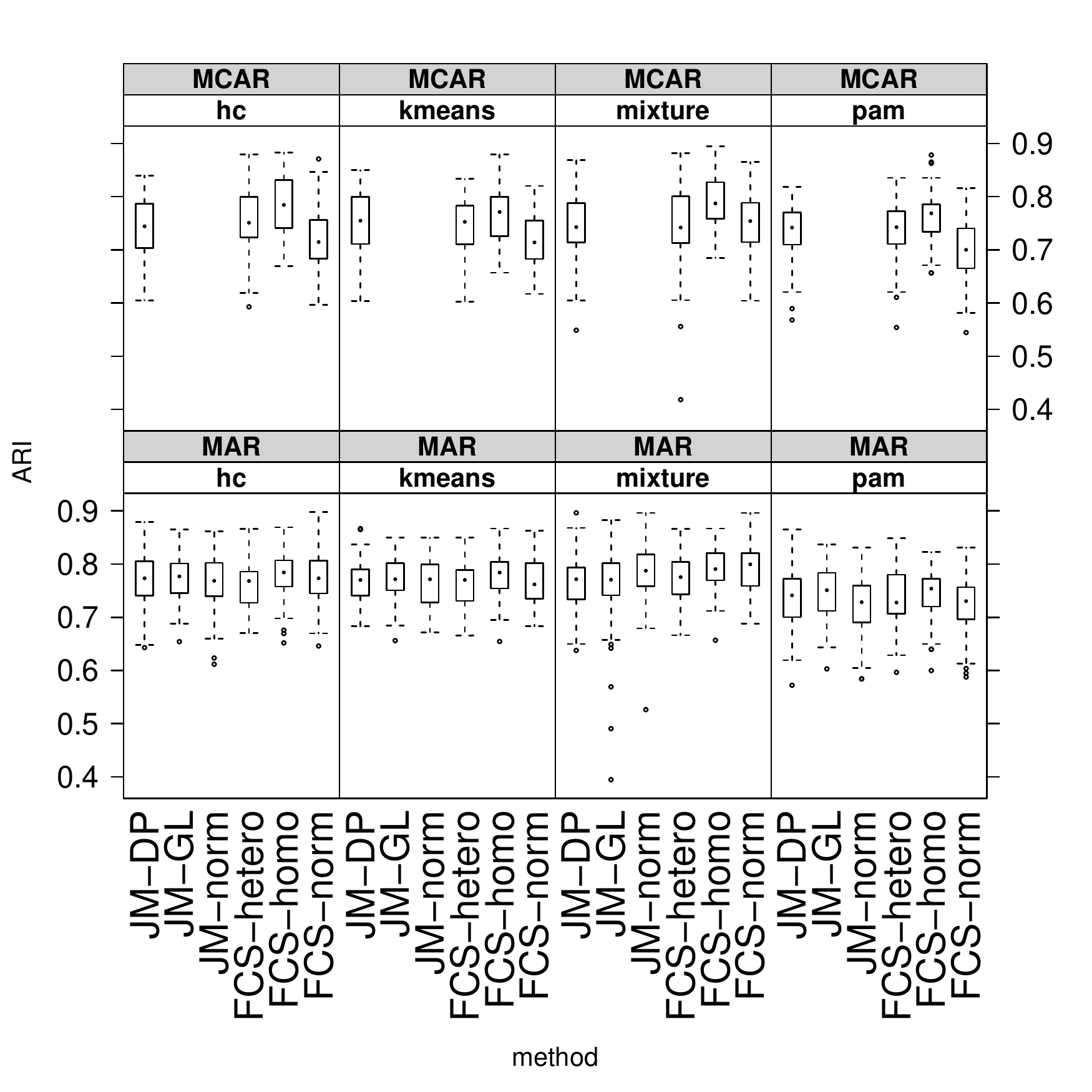}
    \caption{Wine data set: adjusted rand index (ARI) over 100 missing data patterns for various MI methods (FCS-homo, FCS-hetero, FCS-norm, JM-DP, JM-GL and JM-norm) and various cluster analysis methods (mixture, partitioning around medoids, hierarchical clustering, k means) on the incomplete wine data set. Two missing data patterns are investigated: a pattern from a MCAR mechanism (on the top) and one from a MAR mechanism (on the bottom). In both cases, 40\% of values are missing. Results for JM-GL and JM-norm are missing under the MCAR mechanism because of failures of the MI methods.}
   \label{fig:real}
\end{figure}

After multiple imputation ($\Nbtab=20$), cluster analysis is performed using Gaussian mixture model (with constraint of equal volume and shape but variable orientation on the covariance matrices), k-means clustering, partitioning around medoids and hierarchical clustering. ARI are reported in Figure \ref{fig:real}. 

Because the number of variables is relatively large according to the number of observations, some imputation methods (JM-GL and JM-norm) cannot be generally applied. On the contrary, by specifying conditional imputation models, FCS methods are less subject to this issue. Furthermore, according to the ARI, FCS-homo outperforms others imputation models under the MCAR mechanism. Under the MAR mechanism, the advantage of this method is less evident, but it remains among the imputation methods providing the partition with the highest ARI for all cluster analysis methods.

\section{Discussion\label{Section5}}
Addressing missing values by multiple imputation presents several advantages in clustering. It allows assessment of the instability related to missing values and increases partition accuracy as bagging methods. Furthermore, it allows applying classical cluster analysis methods that could not be directly applied with missing values, such as hierarchical clustering. 

 In terms of partition accuracy, the performances of MI methods are mainly related to the missing data pattern: individuals with many missing values cannot be clustered accurately which is more frequent under MAR mechanisms. The data structure is also an important factor to explain the partition accuracy: as in the full data case, the partition is more accurate and more robust to sample variability when clusters are well separated or when the number of individuals is large. 

This study shows cluster analysis requires specific imputation models. Except in the case of two clusters, the accuracy of the partition is improved when the imputation model accounts for underlying class structure. From this point of view, standard MI methods ignoring such a structure (\textit{e.g.} JM-norm, FCS-norm) should be avoided. We identified two relevant JM MI methods handling clustered individuals (JM-GL and JM-DP) and we proposed a new FCS MI method (available in the \textit{clusterMI} R package at the first authors web page). The simulation study highlights JM MI methods performs well 
when data are distributed according to a gaussian mixture model. However, FCS MI can significantly outperform JM MI on more complex designs. Indeed, the flexibility of FCS MI allows specification of each conditional model: to deal with overfitting, by identifying a subset of explanatory variables; to deal with complex relationships between variables, by incorporating interactions effects in the regression models; to deal with semi-continuous variables, by using semi-parametric \citep{Morris14,Gaffert18} or non-parametric imputation models \citep{Doove14}; to deal with outliers, by using robust imputation methods \citep{Templ11}. From this point of view, FCS MI methods are more appealing in practice to deal with non-gaussian distributions.
The gain to use heteroscedastic imputation models (JM-DP, FCS-hetero) is generally limited, even if the covariance varies according to the cluster. Nevertheless, such imputation models can be relevant when missing values are directly related to the cluster structure. Since this relationship can be tricky to identify in practice, we recommend to use heteroscedastic imputation models on data sets where a part of individuals have a large proportion of missing values.
\\

Most of the investigated MI methods required specifying the number of clusters in the data set. In the simulation study, we assumed this number was known, but it can be easily estimated from incomplete data by investigating the instability of the partitions according to the chosen number of clusters \citep{Audigier20}.

Furthermore, these MI methods assume a MAR mechanism. Previous contributions on model-based clustering under MNAR mechanism \cite{Marbac20,Miao16} open the way to propose a JM MI method for a MNAR mechanism, but this topic requires more developments and evaluations.

In addition, we only focus on clustering when data are continuous. A main reason is that categorical variables are more complex in cluster analysis even without missing values. However, the MI methods could handle categorical or mixed data. Concerning JM methods, since JM-GL is dedicated to the imputation of mixed data, its application on mixed data is straightforward; with only categorical variables, it corresponds to the imputation under the log-linear model \citep{Schafer97}; JM-DP does not directly address categorical variables, but it has been also extended to categorical \citep{Si13} and mixed data \citep{Murray15}. Concerning FCS MI methods, dealing with categorical variables is achieved by replacing linear regression models by generalized linear models, like logistic regression for binary variable for instance.

Finally, we focus on dealing with missing values by using MI. However, as stayed in the introduction, some cluster analysis methods have been extended to deal directly with missing values. In a future work, it could be interesting to compare cluster analysis based on MI and cluster analysis avoiding data imputation.

\clearpage
\bibliographystyle{unsrtnat}
\bibliography{references}

\appendix

\section*{Appendix}
\section{Methods}
\subsection{Data Augmentation for JM-GL \label{DAGL}}
\begin{enumerate}
    \item initialize $\bfzeta$ by $\bfzeta^{(0)}=\widehat{\bfzeta}_{ML}$, \textit{i.e.} the maximum likelihood estimator obtained by an EM algorithm
    \item for $\nbiter$ in $\{1,...,\Nbiter\}$
    \begin{description}
    \item[I-step] for each $\nbind$ in $\{1,...,\Nbind\}$
    \begin{enumerate}
        \item draw $w_\nbind^{(\nbiter)}$ from  $p\left(W\vert Z^{obs}=\bfz_\nbind^{obs},\zeta=\bfzeta^{(\nbiter-1)}\right)$
        \item draw ${z_\nbind^{miss}}^{(\nbiter)}$ from $p(Z^{miss}\vert W=\nbgroup_\nbind^{(\nbiter)},Z^{obs}=\bfz_\nbind^{obs},\mu=\bfmu^{(\nbiter-1)},\Sigma=\bfSigma^{(\nbiter-1)})$
    \end{enumerate}
    \item[P-step]~
    \begin{enumerate}
        \item draw $\theta^{(\nbiter)}$ from $p\left(\theta\vert W=\bfw^{(\nbiter)}\right)$
        \item draw $\Sigma^{(\nbiter)}$ from $p\left(\Sigma\vert \theta=\bftheta^{(\nbiter)},Z=\bfZ^{(\nbiter)}\right)$
        \item draw $\mu^{(\nbiter)}$ from $p\left(\mu\vert \theta=\bftheta^{(\nbiter)},\Sigma=\bfSigma^{(\nbiter)},Z=\bfZ^{(\nbiter)},W=\bfw^{(\nbiter)}\right)$
    \end{enumerate}
    \end{description}
\end{enumerate}
with \begin{eqnarray}
p\left(W=\nbgroup\vert Z^{obs},\bfzeta\right)&=&\frac{\theta_\nbgroup exp\left(-\frac{1}{2}\left(\bfz_\nbind^{obs}-\bfmu_\nbgroup^{obs}\right)^t{\bfSigma^{obs}}^{-1}\left(\bfz_\nbind^{obs}-\bfmu_\nbgroup^{obs}\right)\right)}{\sum_{\nbgroup'=1}^\Nbgroup \theta_{\nbgroup'} exp\left(-\frac{1}{2}\left(\bfz_\nbind^{obs}-\bfmu_{\nbgroup'}^{obs}\right)^t{\bfSigma^{obs}}^{-1}\left(\bfz_\nbind^{obs}-\bfmu_{\nbgroup'}^{obs}\right)\right)} \nonumber\\
&\propto&exp\left(\delta_{\nbgroup,\nbind}\right)\label{cond1}
\end{eqnarray}
where $\delta_{\nbgroup,\nbind}={\bfmu_\nbgroup^{obs}}^t{\bfSigma^{obs}}^{-1}z_\nbind^{obs}-\frac{1}{2}{\bfmu_\nbgroup^{obs}}^t{\bfSigma^{obs}}^{-1}\bfmu_\nbgroup^{obs}+log(\theta_\nbgroup)$ is a linear discriminant function based on the reduced information in $\bfz_\nbind^{obs}$ and $\left(\bfmu_\nbgroup^{obs},\bfSigma^{obs}\right)$ denotes the expectation and variance of $Z^{obs}$ given $W$;

\begin{eqnarray}
p(Z^{miss}\vert W=\nbgroup,Z^{obs}=\bfz_\nbind^{obs},\mu=\bfmu^{(\nbiter-1)},\Sigma=\bfSigma^{(\nbiter-1)})
\end{eqnarray}

the conditional distribution of the missing part of $Z$ given the observed one in a specific class. Those distributions are well known in the context of normal data; finally, the P step consists in drawing from the posterior distributions provided in the previous section (Equations \eqref{post1}, \eqref{post2} and \eqref{post3}). 
\subsection{FCS-homoscedastic\label{FCS-homoalgo}}
Denoting $[\bfA,\bfB]$ the matrix obtained by merging two matrices $\bfA$ and $\bfB$ on the rows, $\bfZ_\nbvar^{obs}$ the column $\nbvar$ of $\bfZ$ restricted to its $\Nbind_\nbvar^{obs}$ observed observations (i.e. so that $r_{\nbind\nbvar}=1$ with $\sum_\nbind \left(r_{\nbind\nbvar}\right) =\Nbind_\nbvar^{obs}$), $\bfZ_{-\nbvar}$ the set of continuous variables except column $\nbvar$ and $\bfZ_{-\nbvar}^{obs}$ the set restricted to observed observations on $\bfZ_\nbvar$, the proposed algorithm is as follows:
\begin{enumerate}
    \item Initialization:  
    \begin{itemize}
        \item Using the EM algorithm, find the ML estimate of $\bfzeta$ and initialize $\bfW^{(0)}$ by a draw from the posterior distribution and ${\bfZ^{miss}}^{(0)}$ by a draw from its posterior distribution given $\bfW^{(0)}$
    \end{itemize}
    \item for $\nbiter$ in $\{1,...,\Nbiter\}$
    \begin{enumerate}
    \item $\bfZ\leftarrow \bfZ^{(\nbiter-1)}$
    \item for $\nbvar$ in \{$1,...,\Nbvar\}$
    \begin{enumerate}
        \item \label{eqnquanti}draw $\bfbeta_\nbvar^\star$ and $\sigma_\nbvar^\star$ (the parameters of the linear regression model used to impute the $\nbvar$th variable) from their posterior distribution \citep{VB18}:
        \begin{itemize}
            \item fit the ML estimators $\left(\widehat{\bfbeta}_\nbvar,\widehat{\sigma}_\nbvar\right)$ of the linear regression model $Z_\nbvar^{obs}=\left[\bfU^{(\nbiter-1)}\bfZ_{-\nbvar}\right]^{obs}\bfbeta_\nbvar+\varepsilon$ with $\varepsilon\sim\mathcal{N}\left(0,\sigma^2I_{\Nbind_\nbvar^{obs}}\right)$ 
            \item draw $df$ from $\chi^2(\Nbind_\nbvar^{obs}-\Nbvar-\Nbgroup+1)$ and compute $\sigma_\nbvar^{\star^2}=\left(\hat{\varepsilon}\right)^t\hat{\varepsilon}/df$
            \item draw $\bfbeta_\nbvar^\star$ from \begin{equation} \mathcal{N}\left(\widehat{\bfbeta}_\nbvar,\sigma_\nbvar^{\star^2} \left(\left(\left[\bfU^{(\nbiter-1)}\bfZ_{-\nbvar}^{obs}\right]\right)^t\left[\bfU^{(\nbiter-1)}\bfZ_{-\nbvar}^{obs}\right]\right)^{-1}\right)
            \end{equation}
        \end{itemize}
        \item \label{eqnquanti2}impute the missing values of ${\bfZ_{\nbvar}}$ from  $\mathcal{N}\left(\left[\bfU^{(\nbiter-1)}\bfZ_{-\nbvar}\right]^{miss}\bfbeta_\nbvar^{\star},{\sigma_\nbvar^\star}^2I_{\Nbind_\nbvar^{miss}}\right)$
    \end{enumerate}
    \item $\bfZ^{(\nbiter)}\leftarrow \bfZ$
    \item \label{eqnquali1} by resampling individuals, generate a bootstrap replication $\bfZ^\star$ from $\bfZ^{(\ell)}$ and estimate $\bfzeta$ the set of parameters of a homoscedastic gaussian mixture model. The associated estimate is denoted $\bfzeta^\star$ 
    \item draw $w_\nbind^\star$ (for all $1\leq\nbind\leq\Nbind$) from
    \begin{equation}\label{eqnhomo1}
       P(W=w)\propto exp\left(\left({\bfmu^{\star}_\nbgroup}\right)^t\left(\bfSigma^{\star}\right)^{-1}\bfz_\nbind
    -\frac{1}{2}\left(\bfmu^{\star}_\nbgroup\right)^t\left({\bfSigma^{\star}}\right)^{-1}{\bfmu^{\star}_\nbgroup}+log(\bftheta^{\star}_\nbgroup)\right)
    \end{equation} and derive the disjunctive table $\bfU^{\star}$
    \item draw $\bftheta^{(\nbiter)}$ from $\mathcal{D}\left(\alpha+diag\left(\left(\bfU^{\star}\right)^t\bfU^{\star}\right)\right)$
    \item draw $w^{(\nbiter)}_\nbind$ (for all $1\leq\nbind\leq\Nbind$) from 
    \begin{equation}\label{eqnhomo2}
       P(W=w)\propto exp\left(\left({\bfmu^{\star}_\nbgroup}\right)^t\left(\bfSigma^{\star}\right)^{-1}\bfz_\nbind
    -\frac{1}{2}\left(\bfmu^{\star}_\nbgroup\right)^t\left(\bfSigma^{\star}\right)^{-1}{\bfmu^{\star}_\nbgroup}+log(\bftheta^{(\nbiter)}_\nbgroup)\right)
    \end{equation}
    \end{enumerate}
\end{enumerate}
\subsection{FCS-heteroscedastic\label{FCS-heteroalgo}}
\begin{enumerate}
    \item Initialization:  
    \begin{itemize}
        \item Using the EM algorithm, find the ML estimate of $\bfzeta$ and initialize $\bfW^{(0)}$ by a draw from the posterior distribution and ${\bfZ^{miss}}^{(0)}$ by a draw from its posterior distribution given $\bfW^{(0)}$
    \end{itemize}
    \item for $\nbiter$ in $1..\Nbiter$
    \begin{enumerate}
    \item $\bfZ\leftarrow \bfZ^{(\nbiter-1)}$
    \item for $\nbvar$ in $1...\Nbvar$
    \begin{enumerate}
        \item draw $\bfbeta_\nbvar^\star$ and $\sigma_{\nbvar}^\star=\left(\sigma_{1\nbvar}^\star,...,\sigma_{\Nbgroup\nbvar}^\star\right)$ from their posterior distribution (as proposed in \cite{Resche16} or \cite{vanBuuren11})
        \item impute the missing values of ${\bfz_{\nbind\nbvar}}$ from the linear heteroscedastic regression model corresponding to the cluster $\nbgroup_\nbind$ (for all $1\leq \nbind\leq\Nbind$)
    \end{enumerate}
    \item $\bfZ^{(\nbiter)}\leftarrow \bfZ$
    \item  by resampling individuals, generate a bootstrap replication $\bfZ^\star$ from $\bfZ^{(\ell)}$ and estimate $\bfzeta$ the set of parameters of a heteroscedastic gaussian mixture model. The associated estimate is denoted $\bfzeta^\star$
    \item draw $w_\nbind^\star$ (for all $1\leq\nbind\leq\Nbind$) from
    \begin{equation}
       P(W=w)\propto exp\left(-\frac{1}{2} log\vert{\bfSigma^{\star}_\nbgroup}\vert
    -\frac{1}{2}\left(\bfz_\nbind-\bfmu^{\star}_\nbgroup\right)^t\left({\bfSigma^{\star}_\nbgroup}\right)^{-1}{\left(\bfz_\nbind-\bfmu^{\star}_\nbgroup\right)}+log(\bftheta^{\star}_\nbgroup)\right)
    \end{equation} and derive the disjunctive table $\bfU^{\star}$
    \item draw $\bftheta^{(\nbiter)}$ from $\mathcal{D}\left(\alpha+diag\left(\left(\bfU^{\star}\right)^t\bfU^{\star}\right)\right)$
    \item draw $w^{(\nbiter)}_\nbind$ (for all $1\leq\nbind\leq\Nbind$) from 
    \begin{equation}
       P(W=w)\propto exp\left(-\frac{1}{2} log\vert{\bfSigma^{\star}_\nbgroup}\vert
    -\frac{1}{2}\left(\bfz_\nbind-\bfmu^{\star}_\nbgroup\right)^t\left({\bfSigma^{\star}_\nbgroup}\right)^{-1}{\left(\bfz_\nbind-\bfmu^{\star}_\nbgroup\right)}+log(\bftheta^{(\nbiter)}_\nbgroup)\right)
    \end{equation}
    \end{enumerate}
\end{enumerate}

\section{Simulations}
\subsection{Simulation design}
\begin{sidewaystable}
    \centering
        \caption{Simulation design: values of the parameters used for data generation. Data are generated according to 11 different gaussian mixture models. $\Nbgroup$ denotes the number of components of the mixture, $\Nbind_\nbgroup$ denotes the number of individuals in cluster $\nbgroup$ $(1\leq\nbgroup\leq\Nbgroup)$ and $\left(\bfmu_\nbgroup,\bfSigma_\nbgroup\right)$ denotes the mean and covariance matrix in cluster $\nbgroup$.
    \label{tab:config}}
    $\begin{array}{l|c|cccc|cccc|cccc}
         \text{Configuration}&\Nbgroup&\multicolumn{4}{c|}{\Nbind_\nbgroup}&\multicolumn{4}{c|}{\mu_\nbgroup}&\multicolumn{4}{c}{\Sigma_\nbgroup} \\ \hline
         model-I& 3& 250&250&250&&\mu_a(2)&\mu_b(2)&\mu_c(2)&&\Sigma(0.3)&\Sigma(0.3)&\Sigma(0.3)&\\
         model-II& 3& 250&250&250&&\mu_a(1.5)&\mu_b(1.5)&\mu_c(1.5)&&\Sigma(0.3)&\Sigma(0.3)&\Sigma(0.3)&\\
         model-III& 3& 250&250&250&&\mu_a(2.5)&\mu_b(2.5)&\mu_c(2.5)&&\Sigma(0.3)&\Sigma(0.3)&\Sigma(0.3)&\\
         model-IV& 2& 250&250&&&\mu_a(2)&\mu_b(2)&&&\Sigma(0.3)&\Sigma(0.3)&&\\
         model-V& 4& 250&250&250&250&\mu_a(2)&\mu_b(2)&\mu_c(2)&-\mu_c(2)&\Sigma(0.3)&\Sigma(0.3)&\Sigma(0.3)&\Sigma(0.3)\\
         model-VI& 3& 400&400&400&&\mu_a(2)&\mu_b(2)&\mu_c(2)&&\Sigma(0.3)&\Sigma(0.3)&\Sigma(0.3)&\\
         model-VII& 3& 100&100&100&&\mu_a(2)&\mu_b(2)&\mu_c(2)&&\Sigma(0.3)&\Sigma(0.3)&\Sigma(0.3)\\
         model-VIII& 3& 250&250&100&&\mu_a(2)&\mu_b(2)&\mu_c(2)&&\Sigma(0.3)&\Sigma(0.3)&\Sigma(0.3)&\\
         model-IX& 3& 400&250&250&&\mu_a(2)&\mu_b(2)&\mu_c(2)&&\Sigma(0.3)&\Sigma(0.3)&\Sigma(0.3)&\\
         model-X& 3& 250&250&250&&\mu_a(2)&\mu_b(2)&\mu_c(2)&&I_{4}&\Sigma(0.3)&\Sigma(-0.3)&\\
         model-XI& 3& 250&250&250&&\mu_a(1.5)&\mu_b(1.5)&\mu_c(1.5)&&I_{4}&\Sigma(0.3)&\Sigma(-0.3)&\\
    \end{array}$
    
    $\Sigma(\rho)={\scriptsize{\left(\begin{array}{cc}
    I_4 &  \text{\huge0}\\
    \text{\huge0} & \begin{array}{cccc}1&\rho&\rho&\rho\\
    \rho&1&\rho&\rho\\
    \rho&\rho&1&\rho\\
    \rho&\rho&\rho&1
    \end{array}
\end{array}\right)}}$,  \scriptsize{$\mu_a(\Delta)=(0, 0, 0, 0, \Delta, \Delta, 0, \Delta^2)$,
$\mu_b(\Delta)=(0, 0, 0, 0, -\Delta, -\Delta, -\Delta, 0)$, $\mu_c(\Delta)=(0, 0, 0, 0, -\Delta, \Delta, \Delta,  -\Delta^2)$}
\end{sidewaystable}
\clearpage
\subsection{Results}
\begin{table}[H]
\caption{\label{tab:I}model-I: median adjusted rand index and interquartile range over the $\Nbsim$ generated tables for various multiple imputation methods (JM-DP, JM-GL, JM-norm, FCS-hetero, FCS-homo, FCS-norm), various single imputation method (SI) and various missing data mechanisms (MCAR or MAR with 10\%, 25\% or 40\% of missing values). $\Nbsim<200$ indicates failures of the imputation method. Clustering is performed using finite gaussian mixture models. As benchmark, results before adding missing values using bagging (Full-bagging) or not (Full) are also reported.}
\centering
 \tiny
\begin{tabular}{l|p{.2cm}|p{.3cm}|p{.3cm}|p{.2cm}|p{.3cm}|p{.3cm}|p{.2cm}|p{.3cm}|p{.3cm}|p{.2cm}|p{.3cm}|p{.3cm}|p{.2cm}|p{.3cm}|p{.3cm}|p{.2cm}|p{.3cm}|p{.3cm}}
\hline
\multicolumn{1}{c|}{ } & \multicolumn{9}{c|}{MCAR} & \multicolumn{9}{c}{MAR 1} \\
\cline{2-10} \cline{11-19}
\multicolumn{1}{c|}{ } & \multicolumn{3}{c|}{10\%} & \multicolumn{3}{c|}{25\%} & \multicolumn{3}{c|}{40\%} & \multicolumn{3}{c|}{10\%} & \multicolumn{3}{c|}{25\%} & \multicolumn{3}{c}{40\%} \\
\cline{2-4} \cline{5-7} \cline{8-10} \cline{11-13} \cline{14-16} \cline{17-19}
  & $\Nbsim$ & q50\% & IQ & $\Nbsim$ & q50\% & IQ & $\Nbsim$ & q50\% & IQ & $\Nbsim$ & q50\% & IQ & $\Nbsim$ & q50\% & IQ & $\Nbsim$ & q50\% & IQ\\
\hline
Full-boot & 200 & 0.992 & 0.004 & 200 & 0.992 & 0.004 & 200 & 0.992 & 0.004 & 200 & 0.992 & 0.004 & 200 & 0.992 & 0.004 & 200 & 0.992 & 0.004\\
\hline
Full & 200 & 0.992 & 0.004 & 200 & 0.992 & 0.004 & 200 & 0.992 & 0.004 & 200 & 0.992 & 0.004 & 200 & 0.992 & 0.004 & 200 & 0.992 & 0.004\\
\hline
JM-DP & 200 & 0.980 & 0.012 & 200 & 0.925 & 0.023 & 200 & 0.802 & 0.032 & 200 & 0.948 & 0.023 & 200 & 0.816 & 0.033 & 200 & 0.639 & 0.048\\
\hline
JM-GL & 200 & 0.984 & 0.012 & 200 & 0.929 & 0.023 & 196 & 0.803 & 0.036 & 200 & 0.949 & 0.017 & 199 & 0.817 & 0.034 & 194 & 0.641 & 0.043\\
\hline
JM-norm & 200 & 0.980 & 0.012 & 200 & 0.914 & 0.025 & 200 & 0.770 & 0.038 & 200 & 0.941 & 0.019 & 200 & 0.800 & 0.034 & 200 & 0.610 & 0.045\\
\hline
FCS-hetero & 200 & 0.980 & 0.012 & 200 & 0.918 & 0.027 & 200 & 0.792 & 0.035 & 200 & 0.945 & 0.020 & 200 & 0.813 & 0.036 & 200 & 0.630 & 0.040\\
\hline
FCS-homo & 200 & 0.980 & 0.012 & 200 & 0.921 & 0.019 & 200 & 0.795 & 0.035 & 200 & 0.945 & 0.020 & 200 & 0.813 & 0.036 & 200 & 0.630 & 0.032\\
\hline
FCS-norm & 200 & 0.980 & 0.012 & 200 & 0.910 & 0.023 & 200 & 0.767 & 0.034 & 200 & 0.941 & 0.020 & 200 & 0.802 & 0.037 & 200 & 0.608 & 0.042\\
\hline
JM-DP (SI) & 200 & 0.976 & 0.012 & 200 & 0.906 & 0.030 & 200 & 0.764 & 0.039 & 200 & 0.941 & 0.023 & 200 & 0.799 & 0.036 & 200 & 0.616 & 0.043\\
\hline
JM-GL (SI) & 200 & 0.980 & 0.012 & 200 & 0.914 & 0.020 & 196 & 0.781 & 0.034 & 200 & 0.941 & 0.020 & 199 & 0.806 & 0.031 & 194 & 0.627 & 0.041\\
\hline
JM-norm (SI) & 200 & 0.960 & 0.016 & 200 & 0.857 & 0.027 & 200 & 0.678 & 0.040 & 200 & 0.921 & 0.019 & 200 & 0.749 & 0.033 & 200 & 0.547 & 0.043\\
\hline
FCS-hetero (SI) & 200 & 0.976 & 0.016 & 200 & 0.899 & 0.023 & 200 & 0.753 & 0.036 & 200 & 0.941 & 0.019 & 200 & 0.792 & 0.035 & 200 & 0.610 & 0.038\\
\hline
FCS-homo (SI) & 200 & 0.976 & 0.016 & 200 & 0.901 & 0.022 & 200 & 0.753 & 0.036 & 200 & 0.941 & 0.023 & 200 & 0.793 & 0.033 & 200 & 0.608 & 0.038\\
\hline
FCS-norm (SI) & 200 & 0.960 & 0.020 & 200 & 0.853 & 0.022 & 200 & 0.672 & 0.043 & 200 & 0.918 & 0.023 & 200 & 0.746 & 0.031 & 200 & 0.550 & 0.036\\
\hline
\end{tabular}
\end{table}

\begin{table}[H]
\caption{\label{tab:II}model-II:median adjusted rand index and interquartile range over the $\Nbsim$ generated tables for various multiple imputation methods (JM-DP, JM-GL, JM-norm, FCS-hetero, FCS-homo, FCS-norm), various single imputation method (SI) and various missing data mechanisms (MCAR or MAR with 10\%, 25\% or 40\% of missing values). $\Nbsim<200$ indicates failures of the imputation method. Clustering is performed using finite gaussian mixture models. As benchmark, results before adding missing values using bagging (Full-bagging) or not (Full) are also reported.}
\centering
\tiny
\begin{tabular}{l|p{.2cm}|p{.3cm}|p{.3cm}|p{.2cm}|p{.3cm}|p{.3cm}|p{.2cm}|p{.3cm}|p{.3cm}|p{.2cm}|p{.3cm}|p{.3cm}|p{.2cm}|p{.3cm}|p{.3cm}|p{.2cm}|p{.3cm}|p{.3cm}}
\hline
\multicolumn{1}{c|}{ } & \multicolumn{9}{c|}{MCAR} & \multicolumn{9}{c}{MAR 1} \\
\cline{2-10} \cline{11-19}
\multicolumn{1}{c|}{ } & \multicolumn{3}{c|}{10\%} & \multicolumn{3}{c|}{25\%} & \multicolumn{3}{c|}{40\%} & \multicolumn{3}{c|}{10\%} & \multicolumn{3}{c|}{25\%} & \multicolumn{3}{c}{40\%} \\
\cline{2-4} \cline{5-7} \cline{8-10} \cline{11-13} \cline{14-16} \cline{17-19}
  & $\Nbsim$ & q50\% & IQ & $\Nbsim$ & q50\% & IQ & $\Nbsim$ & q50\% & IQ & $\Nbsim$ & q50\% & IQ & $\Nbsim$ & q50\% & IQ & $\Nbsim$ & q50\% & IQ\\
\hline
Full-boot & 200 & 0.937 & 0.019 & 200 & 0.937 & 0.019 & 200 & 0.937 & 0.019 & 200 & 0.937 & 0.019 & 200 & 0.937 & 0.019 & 200 & 0.937 & 0.019\\
\hline
Full & 200 & 0.937 & 0.019 & 200 & 0.937 & 0.019 & 200 & 0.937 & 0.019 & 200 & 0.937 & 0.019 & 200 & 0.937 & 0.019 & 200 & 0.937 & 0.019\\
\hline
JM-DP & 200 & 0.906 & 0.026 & 200 & 0.813 & 0.032 & 200 & 0.662 & 0.040 & 200 & 0.873 & 0.023 & 200 & 0.723 & 0.036 & 200 & 0.540 & 0.042\\
\hline
JM-GL & 200 & 0.904 & 0.024 & 195 & 0.814 & 0.029 & 192 & 0.665 & 0.039 & 197 & 0.872 & 0.022 & 196 & 0.729 & 0.035 & 191 & 0.542 & 0.040\\
\hline
JM-norm & 200 & 0.899 & 0.026 & 200 & 0.800 & 0.032 & 200 & 0.643 & 0.043 & 200 & 0.865 & 0.023 & 200 & 0.711 & 0.039 & 200 & 0.522 & 0.040\\
\hline
FCS-hetero & 200 & 0.904 & 0.024 & 200 & 0.811 & 0.035 & 200 & 0.659 & 0.039 & 200 & 0.872 & 0.022 & 200 & 0.720 & 0.037 & 200 & 0.534 & 0.040\\
\hline
FCS-homo & 200 & 0.907 & 0.023 & 200 & 0.814 & 0.033 & 200 & 0.670 & 0.042 & 200 & 0.872 & 0.023 & 200 & 0.724 & 0.034 & 200 & 0.543 & 0.039\\
\hline
FCS-norm & 200 & 0.899 & 0.026 & 200 & 0.802 & 0.033 & 200 & 0.643 & 0.046 & 200 & 0.869 & 0.026 & 200 & 0.709 & 0.040 & 200 & 0.522 & 0.044\\
\hline
JM-DP (SI) & 200 & 0.888 & 0.029 & 200 & 0.760 & 0.039 & 200 & 0.586 & 0.044 & 200 & 0.853 & 0.026 & 200 & 0.685 & 0.034 & 200 & 0.492 & 0.042\\
\hline
JM-GL (SI) & 200 & 0.892 & 0.026 & 195 & 0.777 & 0.035 & 192 & 0.611 & 0.044 & 197 & 0.855 & 0.025 & 196 & 0.696 & 0.037 & 191 & 0.504 & 0.040\\
\hline
JM-norm (SI) & 200 & 0.873 & 0.026 & 200 & 0.726 & 0.034 & 200 & 0.533 & 0.048 & 200 & 0.839 & 0.026 & 200 & 0.652 & 0.039 & 200 & 0.453 & 0.047\\
\hline
FCS-hetero (SI) & 200 & 0.891 & 0.026 & 200 & 0.771 & 0.041 & 200 & 0.600 & 0.037 & 200 & 0.856 & 0.025 & 200 & 0.692 & 0.040 & 200 & 0.500 & 0.046\\
\hline
FCS-homo (SI) & 200 & 0.892 & 0.030 & 200 & 0.774 & 0.037 & 200 & 0.607 & 0.038 & 200 & 0.858 & 0.027 & 200 & 0.693 & 0.039 & 200 & 0.507 & 0.040\\
\hline
FCS-norm (SI) & 200 & 0.868 & 0.030 & 200 & 0.719 & 0.041 & 200 & 0.533 & 0.041 & 200 & 0.836 & 0.029 & 200 & 0.646 & 0.045 & 200 & 0.454 & 0.045\\
\hline
\end{tabular}
\end{table}
\begin{table}[H]

\caption{\label{tab:III}model-III: median adjusted rand index and interquartile range over the $\Nbsim$ generated tables for various multiple imputation methods (JM-DP, JM-GL, JM-norm, FCS-hetero, FCS-homo, FCS-norm), various single imputation method (SI) and various missing data mechanisms (MCAR or MAR with 10\%, 25\% or 40\% of missing values). $\Nbsim<200$ indicates failures of the imputation method. Clustering is performed using finite gaussian mixture models. As benchmark, results before adding missing values using bagging (Full-bagging) or not (Full) are also reported.}
\centering
\tiny
\begin{tabular}{l|p{.2cm}|p{.3cm}|p{.3cm}|p{.2cm}|p{.3cm}|p{.3cm}|p{.2cm}|p{.3cm}|p{.3cm}|p{.2cm}|p{.3cm}|p{.3cm}|p{.2cm}|p{.3cm}|p{.3cm}|p{.2cm}|p{.3cm}|p{.3cm}}
\hline
\multicolumn{1}{c|}{ } & \multicolumn{9}{c|}{MCAR} & \multicolumn{9}{c}{MAR 1} \\
\cline{2-10} \cline{11-19}
\multicolumn{1}{c|}{ } & \multicolumn{3}{c|}{10\%} & \multicolumn{3}{c|}{25\%} & \multicolumn{3}{c|}{40\%} & \multicolumn{3}{c|}{10\%} & \multicolumn{3}{c|}{25\%} & \multicolumn{3}{c}{40\%} \\
\cline{2-4} \cline{5-7} \cline{8-10} \cline{11-13} \cline{14-16} \cline{17-19}
  & $\Nbsim$ & q50\% & IQ & $\Nbsim$ & q50\% & IQ & $\Nbsim$ & q50\% & IQ & $\Nbsim$ & q50\% & IQ & $\Nbsim$ & q50\% & IQ & $\Nbsim$ & q50\% & IQ\\
\hline
Full-boot & 200 & 1.000 & 0.000 & 200 & 1.000 & 0.000 & 200 & 1.000 & 0.000 & 200 & 1.000 & 0.000 & 200 & 1.000 & 0.000 & 200 & 1.000 & 0.000\\
\hline
Full & 200 & 1.000 & 0.000 & 200 & 1.000 & 0.000 & 200 & 1.000 & 0.000 & 200 & 1.000 & 0.000 & 200 & 1.000 & 0.000 & 200 & 1.000 & 0.000\\
\hline
JM-DP & 200 & 0.996 & 0.004 & 200 & 0.955 & 0.020 & 200 & 0.842 & 0.029 & 200 & 0.964 & 0.020 & 200 & 0.839 & 0.031 & 200 & 0.666 & 0.040\\
\hline
JM-GL & 200 & 0.996 & 0.008 & 198 & 0.953 & 0.020 & 195 & 0.844 & 0.030 & 200 & 0.964 & 0.016 & 200 & 0.842 & 0.030 & 194 & 0.667 & 0.038\\
\hline
JM-norm & 200 & 0.992 & 0.008 & 200 & 0.941 & 0.027 & 200 & 0.810 & 0.033 & 200 & 0.958 & 0.016 & 200 & 0.826 & 0.032 & 200 & 0.636 & 0.040\\
\hline
FCS-hetero & 200 & 0.992 & 0.004 & 200 & 0.945 & 0.020 & 200 & 0.824 & 0.030 & 200 & 0.960 & 0.016 & 200 & 0.828 & 0.034 & 200 & 0.649 & 0.043\\
\hline
FCS-homo & 200 & 0.992 & 0.004 & 200 & 0.945 & 0.020 & 200 & 0.824 & 0.037 & 200 & 0.960 & 0.016 & 200 & 0.831 & 0.033 & 200 & 0.650 & 0.040\\
\hline
FCS-norm & 200 & 0.992 & 0.008 & 200 & 0.941 & 0.023 & 200 & 0.809 & 0.030 & 200 & 0.956 & 0.020 & 200 & 0.824 & 0.036 & 200 & 0.634 & 0.040\\
\hline
JM-DP (SI) & 200 & 0.992 & 0.008 & 200 & 0.945 & 0.023 & 200 & 0.825 & 0.032 & 200 & 0.960 & 0.017 & 200 & 0.832 & 0.037 & 200 & 0.652 & 0.039\\
\hline
JM-GL (SI) & 200 & 0.996 & 0.004 & 198 & 0.947 & 0.020 & 195 & 0.832 & 0.029 & 200 & 0.964 & 0.012 & 200 & 0.835 & 0.031 & 194 & 0.656 & 0.042\\
\hline
JM-norm (SI) & 200 & 0.982 & 0.012 & 200 & 0.902 & 0.027 & 200 & 0.736 & 0.042 & 200 & 0.941 & 0.019 & 200 & 0.784 & 0.033 & 200 & 0.586 & 0.034\\
\hline
FCS-hetero (SI) & 200 & 0.992 & 0.008 & 200 & 0.929 & 0.023 & 200 & 0.792 & 0.036 & 200 & 0.956 & 0.020 & 200 & 0.817 & 0.035 & 200 & 0.630 & 0.038\\
\hline
FCS-homo (SI) & 200 & 0.992 & 0.008 & 200 & 0.929 & 0.023 & 200 & 0.792 & 0.036 & 200 & 0.953 & 0.019 & 200 & 0.813 & 0.039 & 200 & 0.634 & 0.038\\
\hline
FCS-norm (SI) & 200 & 0.980 & 0.012 & 200 & 0.898 & 0.030 & 200 & 0.729 & 0.034 & 200 & 0.937 & 0.019 & 200 & 0.779 & 0.039 & 200 & 0.581 & 0.040\\
\hline
\end{tabular}
\end{table}
\begin{table}[H]

\caption{\label{tab:IV}model-IV: median adjusted rand index and interquartile range over the $\Nbsim$ generated tables for various multiple imputation methods (JM-DP, JM-GL, JM-norm, FCS-hetero, FCS-homo, FCS-norm), various single imputation method (SI) and various missing data mechanisms (MCAR or MAR with 10\%, 25\% or 40\% of missing values). $\Nbsim<200$ indicates failures of the imputation method. Clustering is performed using finite gaussian mixture models. As benchmark, results before adding missing values using bagging (Full-bagging) or not (Full) are also reported.}
\centering
\tiny
\begin{tabular}{l|p{.2cm}|p{.3cm}|p{.3cm}|p{.2cm}|p{.3cm}|p{.3cm}|p{.2cm}|p{.3cm}|p{.3cm}|p{.2cm}|p{.3cm}|p{.3cm}|p{.2cm}|p{.3cm}|p{.3cm}|p{.2cm}|p{.3cm}|p{.3cm}}
\hline
\multicolumn{1}{c|}{ } & \multicolumn{9}{c|}{MCAR} & \multicolumn{9}{c}{MAR 1} \\
\cline{2-10} \cline{11-19}
\multicolumn{1}{c|}{ } & \multicolumn{3}{c|}{10\%} & \multicolumn{3}{c|}{25\%} & \multicolumn{3}{c|}{40\%} & \multicolumn{3}{c|}{10\%} & \multicolumn{3}{c|}{25\%} & \multicolumn{3}{c}{40\%} \\
\cline{2-4} \cline{5-7} \cline{8-10} \cline{11-13} \cline{14-16} \cline{17-19}
  & $\Nbsim$ & q50\% & IQ & $\Nbsim$ & q50\% & IQ & $\Nbsim$ & q50\% & IQ & $\Nbsim$ & q50\% & IQ & $\Nbsim$ & q50\% & IQ & $\Nbsim$ & q50\% & IQ\\
\hline
Full-boot & 200 & 0.992 & 0.008 & 200 & 0.992 & 0.008 & 200 & 0.992 & 0.008 & 200 & 0.992 & 0.008 & 200 & 0.992 & 0.008 & 200 & 0.992 & 0.008\\
\hline
Full & 200 & 0.992 & 0.008 & 200 & 0.992 & 0.008 & 200 & 0.992 & 0.008 & 200 & 0.992 & 0.008 & 200 & 0.992 & 0.008 & 200 & 0.992 & 0.008\\
\hline
JM-DP & 200 & 0.976 & 0.024 & 200 & 0.952 & 0.024 & 200 & 0.883 & 0.045 & 200 & 0.960 & 0.024 & 200 & 0.868 & 0.047 & 200 & 0.705 & 0.047\\
\hline
JM-GL & 194 & 0.984 & 0.024 & 198 & 0.952 & 0.024 & 196 & 0.883 & 0.040 & 193 & 0.960 & 0.024 & 194 & 0.868 & 0.045 & 194 & 0.712 & 0.054\\
\hline
JM-norm & 200 & 0.984 & 0.024 & 200 & 0.952 & 0.031 & 200 & 0.883 & 0.038 & 200 & 0.956 & 0.024 & 200 & 0.861 & 0.045 & 200 & 0.705 & 0.054\\
\hline
FCS-hetero & 200 & 0.984 & 0.024 & 200 & 0.952 & 0.024 & 200 & 0.891 & 0.045 & 200 & 0.960 & 0.024 & 200 & 0.861 & 0.045 & 200 & 0.705 & 0.054\\
\hline
FCS-homo & 200 & 0.984 & 0.024 & 200 & 0.952 & 0.024 & 200 & 0.883 & 0.040 & 200 & 0.960 & 0.024 & 200 & 0.865 & 0.045 & 200 & 0.705 & 0.061\\
\hline
FCS-norm & 200 & 0.984 & 0.024 & 200 & 0.952 & 0.031 & 200 & 0.883 & 0.038 & 200 & 0.960 & 0.024 & 200 & 0.861 & 0.045 & 200 & 0.705 & 0.049\\
\hline
JM-DP (SI) & 200 & 0.976 & 0.016 & 200 & 0.937 & 0.031 & 200 & 0.853 & 0.046 & 200 & 0.952 & 0.023 & 200 & 0.846 & 0.052 & 200 & 0.682 & 0.053\\
\hline
JM-GL (SI) & 194 & 0.976 & 0.022 & 198 & 0.937 & 0.023 & 196 & 0.861 & 0.052 & 193 & 0.952 & 0.031 & 194 & 0.846 & 0.044 & 194 & 0.685 & 0.064\\
\hline
JM-norm (SI) & 200 & 0.976 & 0.024 & 200 & 0.929 & 0.031 & 200 & 0.831 & 0.044 & 200 & 0.945 & 0.025 & 200 & 0.831 & 0.051 & 200 & 0.665 & 0.059\\
\hline
FCS-hetero (SI) & 200 & 0.976 & 0.018 & 200 & 0.945 & 0.031 & 200 & 0.861 & 0.044 & 200 & 0.952 & 0.024 & 200 & 0.846 & 0.052 & 200 & 0.688 & 0.060\\
\hline
FCS-homo (SI) & 200 & 0.976 & 0.016 & 200 & 0.945 & 0.031 & 200 & 0.865 & 0.039 & 200 & 0.952 & 0.031 & 200 & 0.853 & 0.039 & 200 & 0.692 & 0.060\\
\hline
FCS-norm (SI) & 200 & 0.968 & 0.024 & 200 & 0.921 & 0.031 & 200 & 0.839 & 0.046 & 200 & 0.945 & 0.031 & 200 & 0.831 & 0.044 & 200 & 0.665 & 0.059\\
\hline
\end{tabular}
\end{table}

\begin{table}[H]

\caption{\label{tab:V}model-V: median adjusted rand index and interquartile range over the $\Nbsim$ generated tables for various multiple imputation methods (JM-DP, JM-GL, JM-norm, FCS-hetero, FCS-homo, FCS-norm), various single imputation method (SI) and various missing data mechanisms (MCAR or MAR with 10\%, 25\% or 40\% of missing values). $\Nbsim<200$ indicates failures of the imputation method. Clustering is performed using finite gaussian mixture models. As benchmark, results before adding missing values using bagging (Full-bagging) or not (Full) are also reported.}
\centering
\tiny
\begin{tabular}{l|p{.2cm}|p{.3cm}|p{.3cm}|p{.2cm}|p{.3cm}|p{.3cm}|p{.2cm}|p{.3cm}|p{.3cm}|p{.2cm}|p{.3cm}|p{.3cm}|p{.2cm}|p{.3cm}|p{.3cm}|p{.2cm}|p{.3cm}|p{.3cm}}
\hline
\multicolumn{1}{c|}{ } & \multicolumn{9}{c|}{MCAR} & \multicolumn{9}{c}{MAR 1} \\
\cline{2-10} \cline{11-19}
\multicolumn{1}{c|}{ } & \multicolumn{3}{c|}{10\%} & \multicolumn{3}{c|}{25\%} & \multicolumn{3}{c|}{40\%} & \multicolumn{3}{c|}{10\%} & \multicolumn{3}{c|}{25\%} & \multicolumn{3}{c}{40\%} \\
\cline{2-4} \cline{5-7} \cline{8-10} \cline{11-13} \cline{14-16} \cline{17-19}
  & $\Nbsim$ & q50\% & IQ & $\Nbsim$ & q50\% & IQ & $\Nbsim$ & q50\% & IQ & $\Nbsim$ & q50\% & IQ & $\Nbsim$ & q50\% & IQ & $\Nbsim$ & q50\% & IQ\\
\hline
Full-boot & 200 & 0.976 & 0.008 & 200 & 0.976 & 0.008 & 200 & 0.976 & 0.008 & 200 & 0.976 & 0.008 & 200 & 0.976 & 0.008 & 200 & 0.976 & 0.008\\
\hline
Full & 200 & 0.976 & 0.008 & 200 & 0.976 & 0.008 & 200 & 0.976 & 0.008 & 200 & 0.976 & 0.008 & 200 & 0.976 & 0.008 & 200 & 0.976 & 0.008\\
\hline
JM-DP & 200 & 0.938 & 0.017 & 200 & 0.819 & 0.030 & 200 & 0.645 & 0.028 & 200 & 0.902 & 0.022 & 200 & 0.729 & 0.034 & 200 & 0.522 & 0.031\\
\hline
JM-GL & 174 & 0.938 & 0.016 & 106 & 0.822 & 0.026 & 62 & 0.648 & 0.027 & 177 & 0.901 & 0.020 & 134 & 0.730 & 0.036 & 102 & 0.522 & 0.040\\
\hline
JM-norm & 200 & 0.935 & 0.018 & 200 & 0.812 & 0.031 & 200 & 0.630 & 0.034 & 200 & 0.894 & 0.022 & 200 & 0.717 & 0.034 & 200 & 0.513 & 0.036\\
\hline
FCS-hetero & 200 & 0.935 & 0.020 & 200 & 0.818 & 0.026 & 200 & 0.642 & 0.030 & 200 & 0.897 & 0.022 & 200 & 0.722 & 0.031 & 200 & 0.518 & 0.034\\
\hline
FCS-homo & 200 & 0.937 & 0.016 & 200 & 0.821 & 0.023 & 200 & 0.645 & 0.028 & 200 & 0.900 & 0.020 & 200 & 0.726 & 0.031 & 200 & 0.521 & 0.032\\
\hline
FCS-norm & 200 & 0.935 & 0.015 & 200 & 0.813 & 0.030 & 200 & 0.629 & 0.030 & 200 & 0.893 & 0.020 & 200 & 0.718 & 0.036 & 200 & 0.510 & 0.035\\
\hline
JM-DP (SI) & 200 & 0.923 & 0.019 & 200 & 0.789 & 0.029 & 200 & 0.605 & 0.030 & 200 & 0.887 & 0.021 & 200 & 0.706 & 0.030 & 200 & 0.508 & 0.035\\
\hline
JM-GL (SI) & 174 & 0.928 & 0.015 & 106 & 0.802 & 0.025 & 62 & 0.623 & 0.031 & 177 & 0.892 & 0.019 & 134 & 0.713 & 0.030 & 102 & 0.519 & 0.037\\
\hline
JM-norm (SI) & 200 & 0.905 & 0.022 & 200 & 0.743 & 0.029 & 200 & 0.542 & 0.034 & 200 & 0.869 & 0.022 & 200 & 0.667 & 0.034 & 200 & 0.456 & 0.059\\
\hline
FCS-hetero (SI) & 200 & 0.923 & 0.019 & 200 & 0.786 & 0.031 & 200 & 0.598 & 0.027 & 200 & 0.887 & 0.020 & 200 & 0.699 & 0.032 & 200 & 0.499 & 0.026\\
\hline
FCS-homo (SI) & 200 & 0.925 & 0.015 & 200 & 0.790 & 0.026 & 200 & 0.603 & 0.029 & 200 & 0.887 & 0.024 & 200 & 0.704 & 0.035 & 200 & 0.504 & 0.031\\
\hline
FCS-norm (SI) & 200 & 0.904 & 0.022 & 200 & 0.746 & 0.027 & 200 & 0.545 & 0.032 & 200 & 0.869 & 0.022 & 200 & 0.668 & 0.037 & 200 & 0.455 & 0.053\\
\hline
\end{tabular}
\end{table}
\begin{table}[H]

\caption{\label{tab:VI}model-VI: median adjusted rand index and interquartile range over the $\Nbsim$ generated tables for various multiple imputation methods (JM-DP, JM-GL, JM-norm, FCS-hetero, FCS-homo, FCS-norm), various single imputation method (SI) and various missing data mechanisms (MCAR or MAR with 10\%, 25\% or 40\% of missing values). $\Nbsim<200$ indicates failures of the imputation method. Clustering is performed using finite gaussian mixture models. As benchmark, results before adding missing values using bagging (Full-bagging) or not (Full) are also reported.}
\centering
\tiny
\begin{tabular}{l|p{.2cm}|p{.3cm}|p{.3cm}|p{.2cm}|p{.3cm}|p{.3cm}|p{.2cm}|p{.3cm}|p{.3cm}|p{.2cm}|p{.3cm}|p{.3cm}|p{.2cm}|p{.3cm}|p{.3cm}|p{.2cm}|p{.3cm}|p{.3cm}}
\hline
\multicolumn{1}{c|}{ } & \multicolumn{9}{c|}{MCAR} & \multicolumn{9}{c}{MAR 1} \\
\cline{2-10} \cline{11-19}
\multicolumn{1}{c|}{ } & \multicolumn{3}{c|}{10\%} & \multicolumn{3}{c|}{25\%} & \multicolumn{3}{c|}{40\%} & \multicolumn{3}{c|}{10\%} & \multicolumn{3}{c|}{25\%} & \multicolumn{3}{c}{40\%} \\
\cline{2-4} \cline{5-7} \cline{8-10} \cline{11-13} \cline{14-16} \cline{17-19}
  & $\Nbsim$ & q50\% & IQ & $\Nbsim$ & q50\% & IQ & $\Nbsim$ & q50\% & IQ & $\Nbsim$ & q50\% & IQ & $\Nbsim$ & q50\% & IQ & $\Nbsim$ & q50\% & IQ\\
\hline
Full-boot & 200 & 0.995 & 0.005 & 200 & 0.995 & 0.005 & 200 & 0.995 & 0.005 & 200 & 0.995 & 0.005 & 200 & 0.995 & 0.005 & 200 & 0.995 & 0.005\\
\hline
Full & 200 & 0.995 & 0.005 & 200 & 0.995 & 0.005 & 200 & 0.995 & 0.005 & 200 & 0.995 & 0.005 & 200 & 0.995 & 0.005 & 200 & 0.995 & 0.005\\
\hline
JM-DP & 200 & 0.985 & 0.007 & 200 & 0.929 & 0.017 & 200 & 0.803 & 0.023 & 200 & 0.949 & 0.017 & 200 & 0.814 & 0.028 & 200 & 0.640 & 0.037\\
\hline
JM-GL & 200 & 0.985 & 0.007 & 200 & 0.929 & 0.019 & 195 & 0.805 & 0.030 & 200 & 0.951 & 0.015 & 200 & 0.819 & 0.026 & 197 & 0.639 & 0.037\\
\hline
JM-norm & 200 & 0.980 & 0.010 & 200 & 0.912 & 0.019 & 200 & 0.770 & 0.029 & 200 & 0.943 & 0.015 & 200 & 0.801 & 0.026 & 200 & 0.608 & 0.035\\
\hline
FCS-hetero & 200 & 0.983 & 0.010 & 200 & 0.922 & 0.017 & 200 & 0.796 & 0.025 & 200 & 0.948 & 0.017 & 200 & 0.814 & 0.025 & 200 & 0.630 & 0.032\\
\hline
FCS-homo & 200 & 0.983 & 0.010 & 200 & 0.924 & 0.017 & 200 & 0.795 & 0.027 & 200 & 0.948 & 0.015 & 200 & 0.815 & 0.029 & 200 & 0.633 & 0.034\\
\hline
FCS-norm & 200 & 0.980 & 0.010 & 200 & 0.910 & 0.021 & 200 & 0.772 & 0.030 & 200 & 0.943 & 0.015 & 200 & 0.801 & 0.025 & 200 & 0.611 & 0.034\\
\hline
JM-DP (SI) & 200 & 0.980 & 0.011 & 200 & 0.910 & 0.019 & 200 & 0.773 & 0.032 & 200 & 0.943 & 0.017 & 200 & 0.802 & 0.027 & 200 & 0.620 & 0.038\\
\hline
JM-GL (SI) & 200 & 0.983 & 0.010 & 200 & 0.914 & 0.019 & 195 & 0.781 & 0.028 & 200 & 0.945 & 0.015 & 200 & 0.808 & 0.027 & 197 & 0.623 & 0.032\\
\hline
JM-norm (SI) & 200 & 0.963 & 0.012 & 200 & 0.856 & 0.024 & 200 & 0.676 & 0.034 & 200 & 0.921 & 0.015 & 200 & 0.751 & 0.026 & 200 & 0.549 & 0.036\\
\hline
FCS-hetero (SI) & 200 & 0.975 & 0.007 & 200 & 0.900 & 0.017 & 200 & 0.754 & 0.030 & 200 & 0.939 & 0.015 & 200 & 0.792 & 0.029 & 200 & 0.606 & 0.031\\
\hline
FCS-homo (SI) & 200 & 0.978 & 0.012 & 200 & 0.900 & 0.019 & 200 & 0.755 & 0.028 & 200 & 0.941 & 0.017 & 200 & 0.793 & 0.024 & 200 & 0.605 & 0.036\\
\hline
FCS-norm (SI) & 200 & 0.963 & 0.015 & 200 & 0.853 & 0.021 & 200 & 0.676 & 0.031 & 200 & 0.921 & 0.019 & 200 & 0.750 & 0.033 & 200 & 0.551 & 0.036\\
\hline
\end{tabular}
\end{table}
\begin{table}[H]

\caption{\label{tab:VII}model-VII: median adjusted rand index and interquartile range over the $\Nbsim$ generated tables for various multiple imputation methods (JM-DP, JM-GL, JM-norm, FCS-hetero, FCS-homo, FCS-norm), various single imputation method (SI) and various missing data mechanisms (MCAR or MAR with 10\%, 25\% or 40\% of missing values). $\Nbsim<200$ indicates failures of the imputation method. Clustering is performed using finite gaussian mixture models. As benchmark, results before adding missing values using bagging (Full-bagging) or not (Full) are also reported.}
\centering
\tiny
\begin{tabular}{l|p{.2cm}|p{.3cm}|p{.3cm}|p{.2cm}|p{.3cm}|p{.3cm}|p{.2cm}|p{.3cm}|p{.3cm}|p{.2cm}|p{.3cm}|p{.3cm}|p{.2cm}|p{.3cm}|p{.3cm}|p{.2cm}|p{.3cm}|p{.3cm}}
\hline
\multicolumn{1}{c|}{ } & \multicolumn{9}{c|}{MCAR} & \multicolumn{9}{c}{MAR 1} \\
\cline{2-10} \cline{11-19}
\multicolumn{1}{c|}{ } & \multicolumn{3}{c|}{10\%} & \multicolumn{3}{c|}{25\%} & \multicolumn{3}{c|}{40\%} & \multicolumn{3}{c|}{10\%} & \multicolumn{3}{c|}{25\%} & \multicolumn{3}{c}{40\%} \\
\cline{2-4} \cline{5-7} \cline{8-10} \cline{11-13} \cline{14-16} \cline{17-19}
  & $\Nbsim$ & q50\% & IQ & $\Nbsim$ & q50\% & IQ & $\Nbsim$ & q50\% & IQ & $\Nbsim$ & q50\% & IQ & $\Nbsim$ & q50\% & IQ & $\Nbsim$ & q50\% & IQ\\
\hline
Full-boot & 200 & 0.995 & 0.010 & 200 & 0.995 & 0.010 & 200 & 0.995 & 0.010 & 200 & 0.995 & 0.010 & 200 & 0.995 & 0.010 & 200 & 0.995 & 0.010\\
\hline
Full & 200 & 1.000 & 0.010 & 200 & 1.000 & 0.010 & 200 & 1.000 & 0.010 & 200 & 1.000 & 0.010 & 200 & 1.000 & 0.010 & 200 & 1.000 & 0.010\\
\hline
JM-DP & 200 & 0.980 & 0.020 & 200 & 0.921 & 0.039 & 200 & 0.791 & 0.061 & 200 & 0.950 & 0.039 & 200 & 0.809 & 0.053 & 200 & 0.631 & 0.071\\
\hline
JM-GL & 198 & 0.980 & 0.020 & 192 & 0.921 & 0.038 & 187 & 0.800 & 0.059 & 198 & 0.950 & 0.029 & 196 & 0.817 & 0.047 & 187 & 0.639 & 0.058\\
\hline
JM-norm & 200 & 0.980 & 0.020 & 200 & 0.902 & 0.038 & 200 & 0.764 & 0.070 & 200 & 0.941 & 0.029 & 200 & 0.791 & 0.054 & 200 & 0.614 & 0.066\\
\hline
FCS-hetero & 200 & 0.980 & 0.020 & 200 & 0.912 & 0.038 & 200 & 0.791 & 0.059 & 200 & 0.950 & 0.030 & 200 & 0.809 & 0.048 & 200 & 0.628 & 0.064\\
\hline
FCS-homo & 200 & 0.980 & 0.020 & 200 & 0.921 & 0.038 & 200 & 0.791 & 0.058 & 200 & 0.950 & 0.020 & 200 & 0.809 & 0.055 & 200 & 0.630 & 0.064\\
\hline
FCS-norm & 200 & 0.980 & 0.020 & 200 & 0.902 & 0.038 & 200 & 0.764 & 0.063 & 200 & 0.945 & 0.032 & 200 & 0.799 & 0.054 & 200 & 0.616 & 0.072\\
\hline
JM-DP (SI) & 200 & 0.970 & 0.020 & 200 & 0.892 & 0.048 & 200 & 0.717 & 0.077 & 200 & 0.940 & 0.032 & 200 & 0.791 & 0.062 & 200 & 0.591 & 0.078\\
\hline
JM-GL (SI) & 198 & 0.980 & 0.020 & 192 & 0.912 & 0.038 & 187 & 0.776 & 0.059 & 198 & 0.941 & 0.030 & 196 & 0.800 & 0.054 & 187 & 0.627 & 0.064\\
\hline
JM-norm (SI) & 200 & 0.960 & 0.029 & 200 & 0.846 & 0.054 & 200 & 0.655 & 0.080 & 200 & 0.921 & 0.039 & 200 & 0.746 & 0.060 & 200 & 0.535 & 0.072\\
\hline
FCS-hetero (SI) & 200 & 0.970 & 0.022 & 200 & 0.893 & 0.039 & 200 & 0.739 & 0.062 & 200 & 0.941 & 0.039 & 200 & 0.783 & 0.047 & 200 & 0.606 & 0.062\\
\hline
FCS-homo (SI) & 200 & 0.970 & 0.030 & 200 & 0.902 & 0.047 & 200 & 0.748 & 0.068 & 200 & 0.941 & 0.022 & 200 & 0.790 & 0.054 & 200 & 0.609 & 0.065\\
\hline
FCS-norm (SI) & 200 & 0.960 & 0.020 & 200 & 0.837 & 0.055 & 200 & 0.662 & 0.067 & 200 & 0.921 & 0.039 & 200 & 0.738 & 0.061 & 200 & 0.540 & 0.069\\
\hline
\end{tabular}
\end{table}
\begin{table}[H]

\caption{\label{tab:VIII}model-VIII: median adjusted rand index and interquartile range over the $\Nbsim$ generated tables for various multiple imputation methods (JM-DP, JM-GL, JM-norm, FCS-hetero, FCS-homo, FCS-norm), various single imputation method (SI) and various missing data mechanisms (MCAR or MAR with 10\%, 25\% or 40\% of missing values). $\Nbsim<200$ indicates failures of the imputation method. Clustering is performed using finite gaussian mixture models. As benchmark, results before adding missing values using bagging (Full-bagging) or not (Full) are also reported.}
\centering
\tiny
\begin{tabular}{l|p{.2cm}|p{.3cm}|p{.3cm}|p{.2cm}|p{.3cm}|p{.3cm}|p{.2cm}|p{.3cm}|p{.3cm}|p{.2cm}|p{.3cm}|p{.3cm}|p{.2cm}|p{.3cm}|p{.3cm}|p{.2cm}|p{.3cm}|p{.3cm}}
\hline
\multicolumn{1}{c|}{ } & \multicolumn{9}{c|}{MCAR} & \multicolumn{9}{c}{MAR 1} \\
\cline{2-10} \cline{11-19}
\multicolumn{1}{c|}{ } & \multicolumn{3}{c|}{10\%} & \multicolumn{3}{c|}{25\%} & \multicolumn{3}{c|}{40\%} & \multicolumn{3}{c|}{10\%} & \multicolumn{3}{c|}{25\%} & \multicolumn{3}{c}{40\%} \\
\cline{2-4} \cline{5-7} \cline{8-10} \cline{11-13} \cline{14-16} \cline{17-19}
  & $\Nbsim$ & q50\% & IQ & $\Nbsim$ & q50\% & IQ & $\Nbsim$ & q50\% & IQ & $\Nbsim$ & q50\% & IQ & $\Nbsim$ & q50\% & IQ & $\Nbsim$ & q50\% & IQ\\
\hline
Full-boot & 200 & 0.988 & 0.012 & 200 & 0.988 & 0.012 & 200 & 0.988 & 0.012 & 200 & 0.988 & 0.012 & 200 & 0.988 & 0.012 & 200 & 0.988 & 0.012\\
\hline
Full & 200 & 0.988 & 0.006 & 200 & 0.988 & 0.006 & 200 & 0.988 & 0.006 & 200 & 0.988 & 0.006 & 200 & 0.988 & 0.006 & 200 & 0.988 & 0.006\\
\hline
JM-DP & 200 & 0.980 & 0.014 & 200 & 0.938 & 0.027 & 200 & 0.842 & 0.035 & 200 & 0.951 & 0.018 & 200 & 0.840 & 0.034 & 200 & 0.661 & 0.048\\
\hline
JM-GL & 195 & 0.982 & 0.014 & 197 & 0.942 & 0.024 & 191 & 0.851 & 0.034 & 198 & 0.954 & 0.018 & 195 & 0.841 & 0.032 & 192 & 0.674 & 0.047\\
\hline
JM-norm & 200 & 0.977 & 0.014 & 200 & 0.924 & 0.026 & 200 & 0.810 & 0.041 & 200 & 0.948 & 0.018 & 200 & 0.822 & 0.037 & 200 & 0.632 & 0.067\\
\hline
FCS-hetero & 200 & 0.980 & 0.013 & 200 & 0.935 & 0.025 & 200 & 0.834 & 0.035 & 200 & 0.950 & 0.015 & 200 & 0.835 & 0.033 & 200 & 0.654 & 0.055\\
\hline
FCS-homo & 200 & 0.980 & 0.012 & 200 & 0.935 & 0.022 & 200 & 0.836 & 0.033 & 200 & 0.952 & 0.017 & 200 & 0.839 & 0.034 & 200 & 0.670 & 0.049\\
\hline
FCS-norm & 200 & 0.976 & 0.014 & 200 & 0.924 & 0.026 & 200 & 0.812 & 0.038 & 200 & 0.947 & 0.022 & 200 & 0.825 & 0.034 & 200 & 0.633 & 0.076\\
\hline
JM-DP (SI) & 200 & 0.974 & 0.013 & 200 & 0.915 & 0.028 & 200 & 0.791 & 0.040 & 200 & 0.943 & 0.022 & 200 & 0.813 & 0.039 & 200 & 0.633 & 0.045\\
\hline
JM-GL (SI) & 195 & 0.978 & 0.015 & 197 & 0.924 & 0.023 & 191 & 0.809 & 0.039 & 198 & 0.944 & 0.020 & 195 & 0.821 & 0.030 & 192 & 0.653 & 0.050\\
\hline
JM-norm (SI) & 200 & 0.959 & 0.020 & 200 & 0.864 & 0.035 & 200 & 0.702 & 0.052 & 200 & 0.921 & 0.024 & 200 & 0.769 & 0.037 & 200 & 0.575 & 0.049\\
\hline
FCS-hetero (SI) & 200 & 0.976 & 0.014 & 200 & 0.914 & 0.028 & 200 & 0.796 & 0.040 & 200 & 0.942 & 0.021 & 200 & 0.811 & 0.036 & 200 & 0.635 & 0.045\\
\hline
FCS-homo (SI) & 200 & 0.974 & 0.016 & 200 & 0.917 & 0.024 & 200 & 0.798 & 0.042 & 200 & 0.942 & 0.019 & 200 & 0.818 & 0.033 & 200 & 0.642 & 0.041\\
\hline
FCS-norm (SI) & 200 & 0.961 & 0.020 & 200 & 0.863 & 0.029 & 200 & 0.705 & 0.046 & 200 & 0.924 & 0.024 & 200 & 0.763 & 0.049 & 200 & 0.570 & 0.047\\
\hline
\end{tabular}
\end{table}

\begin{table}[H]
\caption{\label{tab:IX}model-IX: median adjusted rand index and interquartile range over the $\Nbsim$ generated tables for various multiple imputation methods (JM-DP, JM-GL, JM-norm, FCS-hetero, FCS-homo, FCS-norm), various single imputation method (SI) and various missing data mechanisms (MCAR or MAR with 10\%, 25\% or 40\% of missing values). $\Nbsim<200$ indicates failures of the imputation method. Clustering is performed using finite gaussian mixture models. As benchmark, results before adding missing values using bagging (Full-bagging) or not (Full) are also reported.}
\centering
\tiny
\begin{tabular}{l|p{.2cm}|p{.3cm}|p{.3cm}|p{.2cm}|p{.3cm}|p{.3cm}|p{.2cm}|p{.3cm}|p{.3cm}|p{.2cm}|p{.3cm}|p{.3cm}|p{.2cm}|p{.3cm}|p{.3cm}|p{.2cm}|p{.3cm}|p{.3cm}}
\hline
\multicolumn{1}{c|}{ } & \multicolumn{9}{c|}{MCAR} & \multicolumn{9}{c}{MAR 1} \\
\cline{2-10} \cline{11-19}
\multicolumn{1}{c|}{ } & \multicolumn{3}{c|}{10\%} & \multicolumn{3}{c|}{25\%} & \multicolumn{3}{c|}{40\%} & \multicolumn{3}{c|}{10\%} & \multicolumn{3}{c|}{25\%} & \multicolumn{3}{c}{40\%} \\
\cline{2-4} \cline{5-7} \cline{8-10} \cline{11-13} \cline{14-16} \cline{17-19}
  & $\Nbsim$ & q50\% & IQ & $\Nbsim$ & q50\% & IQ & $\Nbsim$ & q50\% & IQ & $\Nbsim$ & q50\% & IQ & $\Nbsim$ & q50\% & IQ & $\Nbsim$ & q50\% & IQ\\
\hline
Full-boot & 200 & 0.993 & 0.007 & 200 & 0.993 & 0.007 & 200 & 0.993 & 0.007 & 200 & 0.993 & 0.007 & 200 & 0.993 & 0.007 & 200 & 0.993 & 0.007\\
\hline
Full & 200 & 0.993 & 0.007 & 200 & 0.993 & 0.007 & 200 & 0.993 & 0.007 & 200 & 0.993 & 0.007 & 200 & 0.993 & 0.007 & 200 & 0.993 & 0.007\\
\hline
JM-DP & 200 & 0.982 & 0.010 & 200 & 0.931 & 0.018 & 200 & 0.815 & 0.031 & 200 & 0.953 & 0.017 & 200 & 0.827 & 0.030 & 200 & 0.653 & 0.042\\
\hline
JM-GL & 200 & 0.983 & 0.008 & 198 & 0.932 & 0.018 & 194 & 0.821 & 0.034 & 198 & 0.953 & 0.016 & 198 & 0.831 & 0.028 & 196 & 0.664 & 0.039\\
\hline
JM-norm & 200 & 0.978 & 0.012 & 200 & 0.914 & 0.023 & 200 & 0.775 & 0.030 & 200 & 0.945 & 0.018 & 200 & 0.808 & 0.029 & 200 & 0.623 & 0.039\\
\hline
FCS-hetero & 200 & 0.981 & 0.010 & 200 & 0.923 & 0.023 & 200 & 0.800 & 0.036 & 200 & 0.946 & 0.018 & 200 & 0.819 & 0.028 & 200 & 0.650 & 0.041\\
\hline
FCS-homo & 200 & 0.982 & 0.010 & 200 & 0.926 & 0.021 & 200 & 0.803 & 0.031 & 200 & 0.948 & 0.018 & 200 & 0.822 & 0.025 & 200 & 0.651 & 0.036\\
\hline
FCS-norm & 200 & 0.979 & 0.013 & 200 & 0.913 & 0.019 & 200 & 0.779 & 0.030 & 200 & 0.944 & 0.017 & 200 & 0.809 & 0.030 & 200 & 0.621 & 0.045\\
\hline
JM-DP (SI) & 200 & 0.977 & 0.013 & 200 & 0.911 & 0.027 & 200 & 0.775 & 0.040 & 200 & 0.943 & 0.019 & 200 & 0.805 & 0.030 & 200 & 0.624 & 0.045\\
\hline
JM-GL (SI) & 200 & 0.979 & 0.013 & 198 & 0.915 & 0.022 & 194 & 0.784 & 0.034 & 198 & 0.946 & 0.019 & 198 & 0.809 & 0.036 & 196 & 0.634 & 0.042\\
\hline
JM-norm (SI) & 200 & 0.960 & 0.016 & 200 & 0.860 & 0.025 & 200 & 0.689 & 0.032 & 200 & 0.923 & 0.020 & 200 & 0.760 & 0.033 & 200 & 0.563 & 0.041\\
\hline
FCS-hetero (SI) & 200 & 0.975 & 0.013 & 200 & 0.901 & 0.023 & 200 & 0.759 & 0.037 & 200 & 0.941 & 0.019 & 200 & 0.798 & 0.033 & 200 & 0.617 & 0.039\\
\hline
FCS-homo (SI) & 200 & 0.976 & 0.011 & 200 & 0.905 & 0.027 & 200 & 0.763 & 0.035 & 200 & 0.940 & 0.020 & 200 & 0.798 & 0.027 & 200 & 0.617 & 0.039\\
\hline
FCS-norm (SI) & 200 & 0.962 & 0.016 & 200 & 0.859 & 0.025 & 200 & 0.687 & 0.037 & 200 & 0.921 & 0.019 & 200 & 0.754 & 0.034 & 200 & 0.561 & 0.037\\
\hline
\end{tabular}
\end{table}

\begin{table}[H]
\caption{\label{tab:X}model-X: median adjusted rand index and interquartile range over the $\Nbsim$ generated tables for various multiple imputation methods (JM-DP, JM-GL, JM-norm, FCS-hetero, FCS-homo, FCS-norm), various single imputation method (SI) and various missing data mechanisms (MCAR or MAR with 10\%, 25\% or 40\% of missing values). $\Nbsim<200$ indicates failures of the imputation method. Clustering is performed using finite gaussian mixture models. As benchmark, results before adding missing values using bagging (Full-bagging) or not (Full) are also reported.}
\centering
\tiny
\begin{tabular}{l|p{.2cm}|p{.3cm}|p{.3cm}|p{.2cm}|p{.3cm}|p{.3cm}|p{.2cm}|p{.3cm}|p{.3cm}|p{.2cm}|p{.3cm}|p{.3cm}|p{.2cm}|p{.3cm}|p{.3cm}|p{.2cm}|p{.3cm}|p{.3cm}}
\hline
\multicolumn{1}{c|}{ } & \multicolumn{9}{c|}{MCAR} & \multicolumn{9}{c}{MAR 1} \\
\cline{2-10} \cline{11-19}
\multicolumn{1}{c|}{ } & \multicolumn{3}{c|}{10\%} & \multicolumn{3}{c|}{25\%} & \multicolumn{3}{c|}{40\%} & \multicolumn{3}{c|}{10\%} & \multicolumn{3}{c|}{25\%} & \multicolumn{3}{c}{40\%} \\
\cline{2-4} \cline{5-7} \cline{8-10} \cline{11-13} \cline{14-16} \cline{17-19}
  & $\Nbsim$ & q50\% & IQ & $\Nbsim$ & q50\% & IQ & $\Nbsim$ & q50\% & IQ & $\Nbsim$ & q50\% & IQ & $\Nbsim$ & q50\% & IQ & $\Nbsim$ & q50\% & IQ\\
\hline
Full-boot & 200 & 1.000 & 0.004 & 200 & 1.000 & 0.004 & 200 & 1.000 & 0.004 & 200 & 1.000 & 0.004 & 200 & 1.000 & 0.004 & 200 & 1.000 & 0.004\\
\hline
Full & 200 & 1.000 & 0.004 & 200 & 1.000 & 0.004 & 200 & 1.000 & 0.004 & 200 & 1.000 & 0.004 & 200 & 1.000 & 0.004 & 200 & 1.000 & 0.004\\
\hline
JM-DP & 200 & 0.984 & 0.008 & 200 & 0.925 & 0.023 & 200 & 0.802 & 0.034 & 200 & 0.952 & 0.016 & 200 & 0.819 & 0.032 & 200 & 0.633 & 0.042\\
\hline
JM-GL & 200 & 0.984 & 0.008 & 198 & 0.925 & 0.019 & 194 & 0.802 & 0.037 & 200 & 0.952 & 0.020 & 200 & 0.821 & 0.032 & 200 & 0.636 & 0.043\\
\hline
JM-norm & 200 & 0.976 & 0.012 & 200 & 0.899 & 0.027 & 200 & 0.753 & 0.043 & 200 & 0.941 & 0.020 & 200 & 0.788 & 0.029 & 200 & 0.594 & 0.046\\
\hline
FCS-hetero & 200 & 0.984 & 0.012 & 200 & 0.918 & 0.027 & 200 & 0.791 & 0.039 & 200 & 0.949 & 0.023 & 200 & 0.809 & 0.037 & 200 & 0.632 & 0.039\\
\hline
FCS-homo & 200 & 0.982 & 0.012 & 200 & 0.918 & 0.027 & 200 & 0.785 & 0.037 & 200 & 0.949 & 0.020 & 200 & 0.804 & 0.030 & 200 & 0.626 & 0.043\\
\hline
FCS-norm & 200 & 0.976 & 0.012 & 200 & 0.902 & 0.023 & 200 & 0.752 & 0.035 & 200 & 0.941 & 0.019 & 200 & 0.788 & 0.028 & 200 & 0.595 & 0.043\\
\hline
JM-DP (SI) & 200 & 0.980 & 0.012 & 200 & 0.910 & 0.027 & 200 & 0.764 & 0.041 & 200 & 0.945 & 0.023 & 200 & 0.799 & 0.036 & 200 & 0.611 & 0.047\\
\hline
JM-GL (SI) & 200 & 0.980 & 0.009 & 198 & 0.910 & 0.023 & 194 & 0.778 & 0.041 & 200 & 0.945 & 0.020 & 200 & 0.805 & 0.032 & 200 & 0.624 & 0.047\\
\hline
JM-norm (SI) & 200 & 0.954 & 0.020 & 200 & 0.828 & 0.033 & 200 & 0.639 & 0.043 & 200 & 0.910 & 0.027 & 200 & 0.732 & 0.045 & 200 & 0.520 & 0.050\\
\hline
FCS-hetero (SI) & 200 & 0.980 & 0.012 & 200 & 0.899 & 0.023 & 200 & 0.744 & 0.037 & 200 & 0.941 & 0.023 & 200 & 0.793 & 0.033 & 200 & 0.603 & 0.047\\
\hline
FCS-homo (SI) & 200 & 0.980 & 0.012 & 200 & 0.899 & 0.027 & 200 & 0.747 & 0.035 & 200 & 0.941 & 0.023 & 200 & 0.792 & 0.033 & 200 & 0.606 & 0.040\\
\hline
FCS-norm (SI) & 200 & 0.952 & 0.020 & 200 & 0.828 & 0.033 & 200 & 0.634 & 0.042 & 200 & 0.914 & 0.030 & 200 & 0.729 & 0.039 & 200 & 0.518 & 0.043\\
\hline
\end{tabular}
\end{table}

\begin{table}[H]
\caption{\label{tab:XI}model-XI: median adjusted rand index and interquartile range over the $\Nbsim$ generated tables for various multiple imputation methods (JM-DP, JM-GL, JM-norm, FCS-hetero, FCS-homo, FCS-norm), various single imputation method (SI) and various missing data mechanisms (MCAR or MAR with 10\%, 25\% or 40\% of missing values). $\Nbsim<200$ indicates failures of the imputation method. Clustering is performed using finite gaussian mixture models. As benchmark, results before adding missing values using bagging (Full-bagging) or not (Full) are also reported.}
\centering
\tiny
\begin{tabular}{l|p{.2cm}|p{.3cm}|p{.3cm}|p{.2cm}|p{.3cm}|p{.3cm}|p{.2cm}|p{.3cm}|p{.3cm}|p{.2cm}|p{.3cm}|p{.3cm}|p{.2cm}|p{.3cm}|p{.3cm}|p{.2cm}|p{.3cm}|p{.3cm}}
\hline
\multicolumn{1}{c|}{ } & \multicolumn{9}{c|}{MCAR} & \multicolumn{9}{c}{MAR 1} \\
\cline{2-10} \cline{11-19}
\multicolumn{1}{c|}{ } & \multicolumn{3}{c|}{10\%} & \multicolumn{3}{c|}{25\%} & \multicolumn{3}{c|}{40\%} & \multicolumn{3}{c|}{10\%} & \multicolumn{3}{c|}{25\%} & \multicolumn{3}{c}{40\%} \\
\cline{2-4} \cline{5-7} \cline{8-10} \cline{11-13} \cline{14-16} \cline{17-19}
  & $\Nbsim$ & q50\% & IQ & $\Nbsim$ & q50\% & IQ & $\Nbsim$ & q50\% & IQ & $\Nbsim$ & q50\% & IQ & $\Nbsim$ & q50\% & IQ & $\Nbsim$ & q50\% & IQ\\
\hline
Full-boot & 200 & 0.956 & 0.017 & 200 & 0.956 & 0.017 & 200 & 0.956 & 0.017 & 200 & 0.956 & 0.017 & 200 & 0.956 & 0.017 & 200 & 0.956 & 0.017\\
\hline
Full & 200 & 0.956 & 0.019 & 200 & 0.956 & 0.019 & 200 & 0.956 & 0.019 & 200 & 0.956 & 0.019 & 200 & 0.956 & 0.019 & 200 & 0.956 & 0.019\\
\hline
JM-DP & 200 & 0.918 & 0.023 & 200 & 0.817 & 0.029 & 200 & 0.665 & 0.037 & 200 & 0.884 & 0.027 & 200 & 0.732 & 0.038 & 200 & 0.545 & 0.045\\
\hline
JM-GL & 198 & 0.914 & 0.026 & 198 & 0.813 & 0.033 & 190 & 0.660 & 0.046 & 200 & 0.887 & 0.024 & 200 & 0.732 & 0.038 & 196 & 0.544 & 0.046\\
\hline
JM-norm & 200 & 0.906 & 0.026 & 200 & 0.792 & 0.036 & 200 & 0.631 & 0.042 & 200 & 0.865 & 0.031 & 200 & 0.699 & 0.044 & 200 & 0.511 & 0.041\\
\hline
FCS-hetero & 200 & 0.918 & 0.027 & 200 & 0.813 & 0.032 & 200 & 0.661 & 0.045 & 200 & 0.884 & 0.030 & 200 & 0.732 & 0.039 & 200 & 0.545 & 0.047\\
\hline
FCS-homo & 200 & 0.914 & 0.026 & 200 & 0.813 & 0.036 & 200 & 0.662 & 0.040 & 200 & 0.884 & 0.030 & 200 & 0.731 & 0.041 & 200 & 0.544 & 0.044\\
\hline
FCS-norm & 200 & 0.906 & 0.026 & 200 & 0.792 & 0.035 & 200 & 0.627 & 0.041 & 200 & 0.868 & 0.029 & 200 & 0.704 & 0.044 & 200 & 0.509 & 0.047\\
\hline
JM-DP (SI) & 200 & 0.895 & 0.026 & 200 & 0.764 & 0.038 & 200 & 0.588 & 0.043 & 200 & 0.868 & 0.026 & 200 & 0.692 & 0.041 & 200 & 0.497 & 0.045\\
\hline
JM-GL (SI) & 198 & 0.897 & 0.023 & 198 & 0.774 & 0.035 & 190 & 0.606 & 0.040 & 200 & 0.871 & 0.026 & 200 & 0.698 & 0.040 & 196 & 0.506 & 0.050\\
\hline
JM-norm (SI) & 200 & 0.863 & 0.031 & 200 & 0.695 & 0.050 & 200 & 0.490 & 0.055 & 200 & 0.830 & 0.036 & 200 & 0.625 & 0.052 & 200 & 0.417 & 0.052\\
\hline
FCS-hetero (SI) & 200 & 0.895 & 0.027 & 200 & 0.767 & 0.038 & 200 & 0.589 & 0.050 & 200 & 0.865 & 0.026 & 200 & 0.691 & 0.039 & 200 & 0.502 & 0.045\\
\hline
FCS-homo (SI) & 200 & 0.895 & 0.026 & 200 & 0.774 & 0.043 & 200 & 0.605 & 0.044 & 200 & 0.868 & 0.030 & 200 & 0.695 & 0.040 & 200 & 0.512 & 0.042\\
\hline
FCS-norm (SI) & 200 & 0.861 & 0.023 & 200 & 0.691 & 0.046 & 200 & 0.485 & 0.044 & 200 & 0.827 & 0.030 & 200 & 0.625 & 0.045 & 200 & 0.421 & 0.058\\
\hline
\end{tabular}
\end{table}

\end{document}